\documentclass[structabstract]{aa}  
\usepackage{graphicx}
\usepackage{txfonts}
\def\lsim{\lower.5ex\hbox{$\; \buildrel < \over \sim \;$}}
\def\gsim{\lower.5ex\hbox{$\; \buildrel > \over \sim \;$}}
\begin{document}
\title{Radio emission and nonlinear diffusive shock acceleration of cosmic rays in the supernova SN~1993J}

\author{V. Tatischeff}

\offprints{V. Tatischeff}

\institute{Centre de Spectrom\'etrie Nucl\'eaire et de Spectrom\'etrie de 
Masse, CNRS/IN2P3 and Univ Paris-Sud, F-91405 Orsay, France\thanks{Permanent address.} \\
and Institut de Ci\`encies de l'Espai (CSIC-IEEC), Campus UAB, 
Fac. Ci\`encies, 08193 Bellaterra, Barcelona, Spain
\\
              \email{Vincent.Tatischeff@csnsm.in2p3.fr}
             }

\date{Received ...; accepted ...}

   \abstract {}
{The extensive observations of the supernova SN~1993J at radio wavelengths make this object a unique target for the study of particle acceleration in a supernova shock.}
{To describe the radio synchrotron emission we use a model that couples a semianalytic description of nonlinear diffusive shock acceleration with self-similar solutions for the hydrodynamics of the supernova expansion. The synchrotron emission, which is assumed to be produced by relativistic electrons propagating in the postshock plasma, is worked out from radiative transfer calculations that include the process of synchrotron self-absorption. The model is applied to explain the morphology of the radio emission deduced from high-resolution VLBI imaging observations and the measured time evolution of the total flux density at six frequencies.}
{Both the light curves and the morphology of the radio emission indicate that the magnetic field was strongly amplified in the blast wave region shortly after the explosion, possibly via the nonresonant regime of the cosmic-ray streaming instability operating in the shock precursor. The amplified magnetic field immediately upstream from the subshock is determined to be $B_u \approx 50 (t/1 {\rm~day})^{-1}$~G. The turbulent magnetic field was not damped behind the shock but carried along by the plasma flow in the downstream region. Cosmic-ray protons were efficiently produced by diffusive shock acceleration at the blast wave. We find that during the first $\sim$8.5~years after the explosion, about 19\% of the total energy processed by the forward shock was converted to cosmic-ray energy. However, the shock remained weakly modified by the cosmic-ray pressure. The high magnetic field amplification implies that protons were rapidly accelerated to energies well above 10$^{15}$~eV. The results obtained for this supernova support the scenario that massive stars exploding into their former stellar wind are a major source of Galactic cosmic-rays of energies above $\sim$10$^{15}$~eV. We also calculate the flux from SN~1993J of gamma-rays arising from collisions of accelerated cosmic rays with ambient material and the result suggests that type II supernovae could be detected in $\pi^0$-decay gamma-rays with the 
\emph{Fermi} Gamma-ray Space Telescope out to a maximum distance of only $\sim$1~Mpc.}
{}


\titlerunning{Radio emission and nonlinear diffusive shock acceleration in SN~1993J}

   \maketitle
%

\section{Introduction}

Galactic cosmic rays are widely believed to be accelerated in expanding shock waves initiated by supernova (SN) explosions, at least up to the "knee" energy of the cosmic-ray spectrum at $\sim$10$^{15}$~eV and possibly up to the "ankle" at $\sim$10$^{18}$~eV (Axford \cite{axf94}). The theory of diffusive shock  acceleration (DSA) of cosmic rays is well established (see, e.g., Jones \& Ellison \cite{jon91}; Malkov \& Drury \cite{mal01} for reviews), but two fundamental questions remain partly unanswered by the theory: what is the maximum kinetic energy achieved by particles accelerated in SN shocks and what is the acceleration efficiency, i.e. the fraction of the total SN energy converted to cosmic-ray energy? 

The maximum cosmic-ray energy attainable in a SN shock scales as the product of the shock size and the strength of the turbulent magnetic field in the acceleration region (e.g. Marcowith et al. \cite{mar06}). Multiwavelength observations of young shell-type supernova remnants (SNRs) provide evidence that the magnetic field in the blast wave region is much higher than the interstellar magnetic field (e.g. V\"olk et al. \cite{vol05}; Cassam-Chena\"i et al. \cite{cas07}; Uchiyama et al. \cite{uch07} and references therein). A possible explanation for these observations is that the ambient magnetic field is amplified by the diffusive streaming of accelerated particles in the upstream region of the shock, which could cause a plasma instability generating large-amplitude magnetic turbulence (Bell \& Lucek \cite{bel01}; Amato \& Blasi \cite{ama06}; Vladimirov et al. \cite{vla06}). Magnetic field amplification via this nonlinear process could facilitate the acceleration of protons in SNRs up to $\sim$10$^{15}$~eV (e.g. Parizot et al. \cite{par06}). 

The acceleration efficiency depends on the diffusive transport of energetic particles at both sides of the shock in the self-generated turbulence (e.g. Malkov \& Drury \cite{mal01}). The DSA theory cannot accurately predict at the present time how many particles are injected into the acceleration process as a function of the shock parameters. Observations of SNRs in high-energy gamma-rays can provide valuable information on the efficiency of cosmic-ray acceleration in these objects. But it remains unclear whether the emission detected from several SNRs with ground-based atmospheric Cherenkov telescopes is due mainly to pion decay following hadronic collisions of accelerated ions or to inverse Compton scattering of ambient photons by accelerated electrons (e.g. Morlino et al. \cite{mor08}). 

However, theory predicts that efficient acceleration of cosmic-ray ions (mainly protons) modifies the shock structure with respect to the case with no acceleration (see, e.g., Berezhko \& Ellison \cite{ber99}). In particular, the total compression ratio of a cosmic-ray-modified shock can be much higher than that of a test-particle shock (i.e. when the accelerated particles have no influence on the shock structure), because relativistic particles produce less pressure for a given energy density than do nonrelativistic particles. In addition, the energy loss due to escape of accelerated cosmic rays from the shock system can further increase the compressibility of the shocked gas (see Decourchelle et al. \cite{dec00}). Furthermore, since the energy going into relativistic particles is drawn from the shocked-heated thermal population, the postshock temperature of a cosmic-ray-modified shock can be much lower than the test-particle value. Observations of these nonlinear effects (Hughes et al. \cite{hug00}; Decourchelle \cite{dec05}; Warren et al. \cite{war05}) provide indirect evidence for the efficient acceleration of cosmic-ray protons (which carry most of the total nonthermal particle pressure) in shock waves of young SNRs. 

In this paper, we study the production of cosmic-rays by nonlinear DSA and the associated magnetic field amplification in a very young SN shock, which expands in a relatively dense stellar wind lost by the progenitor star prior to explosion. We use radio monitoring observations of SN~1993J conducted with the Very Large Array and several other radio telescopes since the SN outburst (see Weiler et al. \cite{wei07}; Bartel et al. \cite{bar07} and references therein). 
The radio emission from SNe is thought to be synchrotron radiation from relativistic electrons of energies $<1$~GeV accelerated at the expanding blast wave (Chevalier \cite{che82b}; Fransson \& Bj\"ornsson \cite{fra98}). The key motivation for the present study is that the radiating electrons can be influenced by the presence of otherwise unseen shock-accelerated protons. In particular, the theory of nonlinear DSA predicts that the energy distribution of nonthermal electrons below 1~GeV steepens with increasing efficiency of proton acceleration (Ellison et al. \cite{ell00}). 

We chose SN~1993J because it is one of the brightest radio SNe ever detected and has already been the subject of numerous very useful observational and theoretical studies (see, e.g., for the radio emission of SN~1993J Bartel et al. \cite{bar94,bar00,bar07}; Marcaide et al. \cite{mar94,mar95,mar97}; Van Dyk et al. \cite{van94}; Fransson et al. \cite{fra96}; Fransson \& Bj\"ornsson \cite{fra98}; P\'erez-Torres et al. \cite{per01}; Mioduszewski et al. \cite{mio01}, Bietenholz et al. \cite{bie03}, Weiler et al. \cite{wei07} and references therein). 

The model of the present paper is largely based on the work of Cassam-Chena\"i et al. (\cite{cas05}; see also Decourchelle et al. \cite{dec00}; Ellison \& Cassam-Chena\"i \cite{ell05}) on the morphology of synchrotron emission in Galactic SNRs. These authors have developed a model to study radio and X-ray images of SNRs undergoing efficient cosmic-ray production, that couples a semianalytic description of nonlinear DSA with self-similar solutions for the hydrodynamics of the SN expansion. The model has recently been applied to the remnants of Tycho's SN (Cassam-Chena\"i et al. \cite{cas07}) and SN~1006 (Cassam-Chena\"i et al. \cite{cas08}).  

Duffy et al. (\cite{duf95}) have studied the radio emission from SN~1987A by considering a two-fluid system consisting of a cosmic-ray gas and a thermal plasma to calculate the structure of the blast wave. In their model, however, the efficiency of cosmic-ray acceleration is not deduced from the radio data, but assumed to be similar to that required in SNRs to explain the observed flux of Galactic cosmic rays. Another difference with the present work is that the turbulent magnetic field in the shock precursor is not assumed to be amplified by the DSA process, but is taken by these authors to be of the same order as the ordered field in the wind of the progenitor star. 

A preliminary account of the present work has been given elsewhere (Tatischeff \cite{tat08}) and all of the present results supersede those published earlier.

\section{Model}

\subsection{Forward shock expansion}

The type IIb SN~1993J was discovered in the galaxy M81 by Garcia (Ripero et al. \cite{rip93}) on 1993 March 28, shortly after shock breakout (Wheeler et al. \cite{whe93}). Very long baseline interferometry (VLBI) observations (see, e.g., Marcaide et al. \cite{mar97}; Bietenholz et al. \cite{bie03}) revealed a decelerating expansion of a shell-like radio source. The radio emission is presumably produced between the forward shock propagating into the circumstellar medium (CSM) and the reverse shock running into the SN ejecta (e.g. Bartel et al. \cite{bar07}). Marcaide et al. (\cite{mar97}) found the outer angular radius of the radio shell to evolve as a function of time $t$ after explosion with the power law $\theta_o \propto t^m$, where the deceleration parameter $m=0.86\pm0.02$. More recently, Weiler et al. (\cite{wei07} and references therein) reported that the angular expansion of SN~1993J up to day 1500 after outburst can be expressed as $\theta=6.2 \times (t/1~{\rm day})^m$~$\mu$as with $m=0.845\pm0.005$; here $\theta$ is by definition the angular radius of the circle that encompasses half of the total radio flux density to better than 20\%.

These results are consistent with the standard, analytical model for the expansion of a SN into a CSM (Chevalier \cite{che82a,che83}; Nadyozhin \cite{nad85}). This model assumes power-law density profiles for both the outer SN ejecta, $\rho_{\rm ej} = C_2 t^{n-3} R^{-n}$ (with $n>5$), and the CSM, $\rho_{\rm CSM} = C_1 R^{-s}$ (with $s<3$), where $C_2$ and $C_1$ are constants. The SN expansion is then found to be self-similar (i.e. the structure of the interaction region between the forward and reverse shocks remains constant in time except for a scaling factor) and the deceleration parameter $m=(n-3)/(n-s)$. For a standard wind density profile with $s=2$ (see below), the deceleration reported by Weiler et al. (\cite{wei07}) corresponds to $n \approx 8.5$, in fair agreement with numerical computations of SN explosions (e.g. Arnett \cite{arn88}). 

\begin{figure}
   \centering
   \includegraphics[width=0.45\textwidth]{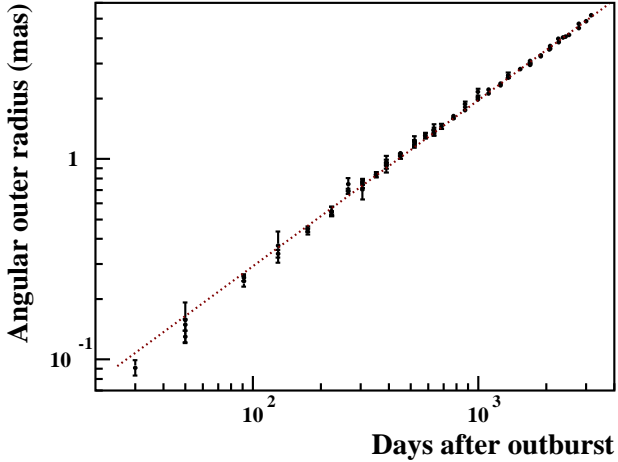}
      \caption{Time evolution of the outer angular radius of the shell-like radio emission from SN~1993J. The data were obtained by Bartel et al. (\cite{bar02}) from observations at 1.7, 2.3, 5.0, 8.4, 14.8, and 22.2~GHz. The dotted line is a least-squares power-law fit to the data (Eq.~\ref{eq1}). 
              }
         \label{fig1}
   \end{figure}

However, Bartel et al. (\cite{bar00,bar02}) found significant changes with time of the parameter $m$, indicating deviations from a self-similar expansion. These authors determined the outer angular radius $\theta_o$ as a function of time by consistently fitting to the two-dimensional radio images observed at 34 epochs between 1993 and 2001 the projection of a three-dimensional spherical shell of uniform volume emissivity. They fixed the ratio of the outer to inner angular radius (as expected from a self-similar expansion) at $\theta_o/\theta_i=1.25$. By performing a least-squares power-law fit to the values of $\theta_o$ thus determined (see Fig.~\ref{fig1}), they obtained a minimum reduced $\chi^2$ of $\chi^2_\nu=1.8$ for 64 degrees of freedom, indicating that the null hypothesis of a self-similar expansion can be rejected.

As discussed by Bartel et al. (\cite{bar02}), this result strongly depends on the systematic errors associated with the model of uniform emissivity in the spherical shell. The overall systematic uncertainty was estimated to range from 3\% at early epochs to 1\% at later epochs and to dominate the statistical error in most of the cases. We show below (Sect.~3.1) that the assumption of uniform emissivity is questionable, because radio-emitting electrons accelerated at the forward shock are expected to lose most of their kinetic energy by radiative losses before reaching the contact discontinuity between shocked ejecta and shocked CSM (see also Fransson \& Bj\"ornsson \cite{fra98}). Furthermore, the radial emissivity profile is expected to vary with time and radio frequency\footnote{This frequency dependence may explain why the outer radii obtained by Bartel et al. (\cite{bar02}) at 2.3 and 1.7~GHz tend to be larger by up to 5\% than those measured by the same authors at higher frequencies.}. Although the systematic uncertainties were carefully studied by Bartel et al. (\cite{bar02}), we note that a moderate increase of the errors would make the expansion compatible with the self-similar assumption. For example, by setting a lower limit of 3\% on the uncertainties in the outer radii measured by Bartel et al., we get from a least-squares power-law fit to the data $\chi^2_\nu=1.25$ and an associated probability of chance coincidence of 8\%. Thus, the null hypothesis of a self-similar expansion could not anymore be rejected at the usual significance level of 5\%. 

For simplicity, I shall use the self-similar solution to model the hydrodynamic evolution of the SNR (Sect.~2.5). The radius of the forward shock, $R_s=\theta_o D$ where the source distance $D=3.63 \pm 0.34$~Mpc (Freedman et al. \cite{fre94}), is estimated from a power-law fit to the data of Bartel et al. (\cite{bar02}) at all frequencies (Fig.~\ref{fig1}), which gives
\begin{equation}
\theta_o = (0.292 \pm 0.004) \bigg({t \over 100{\rm~days}}\bigg)^{0.829 \pm 0.005}~~{\rm mas}.
\label{eq1}
\end{equation}
Thus, we have $R_s=R_0 (t/1~{\rm day})^m$ with $R_0=3.49\times 10^{14}$~cm and $m=0.83$. The forward shock velocity is then
\begin{equation}
V_s = {dR_s \over dt} = V_0 \bigg({t \over 1{\rm~day}}\bigg)^{m-1}~~{\rm with}~V_0 = 3.35 \times 10^4~{\rm km~s}^{-1}.
\label{eq2}
\end{equation}

In the self-similar model, the forward shock radius is given by (e.g. Chevalier \cite{che83})
\begin{equation}
R_s = K_s \bigg({C_2 \over C_1}\bigg)^{1/(n-s)}t^{(n-3)/(n-s)}~,
\label{eq2p}
\end{equation}
where the constant $K_s$ depends on $s$, $n$, and the adiabatic index of the shocked gas $\gamma_g$. For $s=2$, $n=7.88$, corresponding to $m=(n-3)/(n-s)=0.83$, and $\gamma_g=5/3$ (see below), we have $K_s=0.93$. The  parameter $C_1$ that fixes the CSM density can be estimated from the radio emission. Anticipating the results presented in Sect.~3, the best-fit model to the radio light curves gives $C_1=1.9 \times 10^{14}$~g~cm$^{-1}$. Comparison of Eq.~(\ref{eq2p}) with the observed SN expansion then provides $C_2$, which in turn can be expressed in terms of the explosion energy and ejected mass (e.g. Nadyozhin \cite{nad85}; Decourchelle \& Ballet \cite{dec94}). We finally obtain
\begin{equation}
M_{\rm ej} = 2.2 f_{\rm ej} E_{51}^{1.7}~M_\odot~,
\label{eq2t}
\end{equation}
where $f_{\rm ej}$ is a parameter of order unity that depends on the velocity distribution of the inner ejecta and $E_{51}$ is the kinetic energy of the explosion in units of 10$^{51}$~erg. This result is consistent with the ejected mass estimated from the optical light curve of SN~1993J, $1.9<M_{\rm ej}<3.5~M_\odot$ (Young et al. \cite{you95}). However, numerical simulations of SN explosions show that the density profile of the outer ejecta can be more complicated than a power law (see, e.g., Iwamoto et al. \cite{iwa97}) and one should bear in mind that the self-similar solution is a strong simplification of the real hydrodynamic evolution. 

\subsection{Free-free absorption of the radio emission in the circumstellar 
medium}

\begin{figure*}
   \centering
   \includegraphics[width=0.65\textwidth]{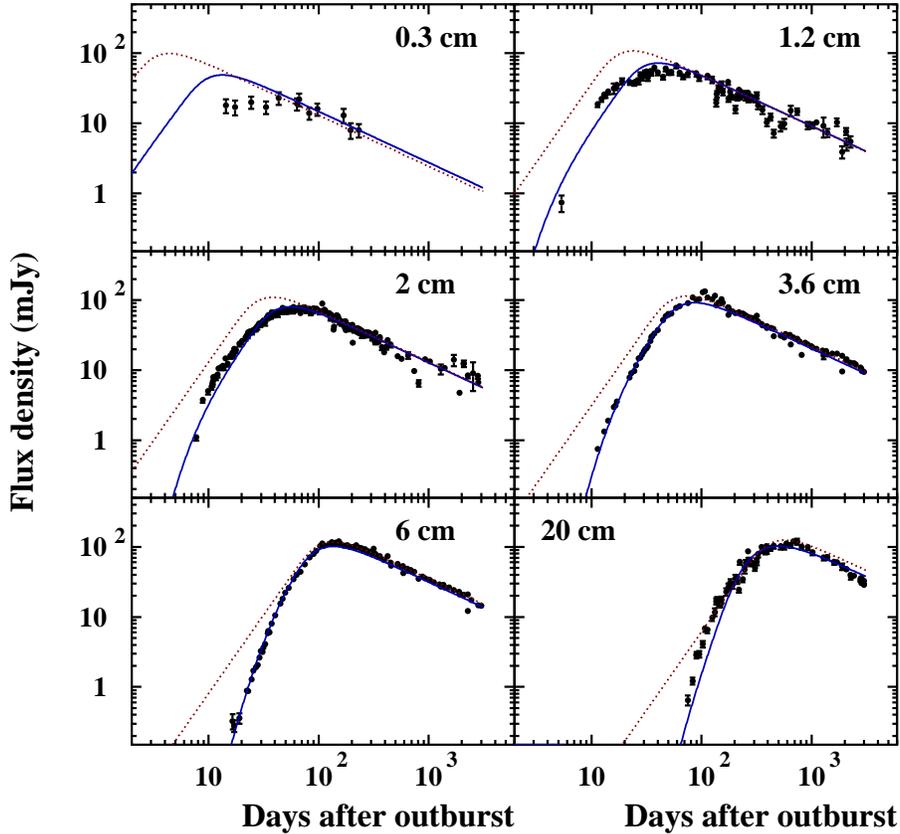}
      \caption{Radio light curves for SN~1993J at 0.3, 1.2, 2, 3.6, 6, and 20~cm. The data are from Weiler et al. (\cite{wei07}) and references therein. The solid lines represent the best-fit semi-empirical model of these authors. The dotted lines show a synchrotron self-absorption (SSA) model in which the SSA optical depth and the unabsorbed flux density were obtained from fits to radio spectra taken between day 75 and day 923 after outburst (see Sect.~2.3). The difference between the latter model and the data before day 100 is partly due to free-free absorption of the radio emission in the CSM. 
               } 
         \label{fig2}
   \end{figure*}

Figure~\ref{fig2} shows a set of light curves measured for SN~1993J at 0.3~cm (85--110 GHz), 1.2~cm (22.5~GHz), 2~cm (14.9~GHz), 3.6~cm (8.4~GHz), 6~cm (4.9~GHz), and 20~cm (1.4~GHz). We see that at each wavelength the flux density first rapidly increases and then declines more slowly as a power in time (the data at 0.3~cm do not allow to clearly identify this behavior). The radio emission was observed to suddenly decline after day $\sim$3100 (not shown in Fig.~\ref{fig2}), which is interpreted in terms of an abrupt decrease of the CSM density at radial distance from the progenitor $R_{\rm out} \sim3\times10^{17}$~cm (Weiler et al. \cite{wei07}). This is presumably the outer limit of the dense cocoon which was established by a high mass loss from the red supergiant progenitor of the SN for $\sim$10$^4$~years before explosion. 

We see in Figure~\ref{fig2} that the maximum intensity is reached first at lower wavelengths and later at higher wavelengths, which is characteristic of absorption processes. For SN~1993J, both free-free absorption (FFA) in the CSM and synchrotron self-absorption (SSA) are important (Chevalier \cite{che98}; Fransson \& Bj\"ornsson \cite{fra98}; Weiler et al. \cite{wei07}). The Razin effect can be excluded (Fransson \& Bj\"ornsson \cite{fra98}). FFA is produced in the stellar wind that has been heated and ionized by radiation from the shock breakout. If the circumstellar gas is homogeneous and of uniform temperature $T_{\rm CSM}$, the radio emission produced behind the forward shock is typically attenuated by a factor $\exp(-\tau_{\rm CSM}^{\rm homog})$, where the optical depth at a given frequency $\nu$ satisfies (Weiler et al. \cite{wei86})
\begin{eqnarray}
\tau_{\rm CSM}^{\rm homog} & \propto & \int_{R_s}^{\infty} \rho_{\rm CSM}^2 T_{\rm CSM}^{-1.35} \nu^{-2.1} dR \propto T_{\rm CSM}^{-1.35} \nu^{-2.1} R_s^{1-2s}  \nonumber \\ 
& \propto & T_{\rm CSM}^{-1.35} \nu^{-2.1} t^{m(1-2s)}~.
\label{eq3}
\end{eqnarray}
From a fit to early data using the parametrized model of Weiler et al. (\cite{wei86,wei02}; see Appendix~A), Van Dyk et al. (\cite{van94}) found for the time dependance of the optical depth $\delta=m(1-2s) \approx -2$. They concluded that the density profile of the CSM must be significantly flatter, $s \sim 1.5$, than that, $s=2$, produced by a constant mass-loss rate and constant-velocity stellar wind. They interpreted this result in terms of a steady decrease of the mass-loss rate of the SN progenitor star prior to explosion. This conclusion was later confirmed by Fransson et al. (\cite{fra96}). But in both studies SSA was not taken into account.

Fransson \& Bj\"ornsson (\cite{fra98}) performed the most detailed modeling of the radio emission from SN~1993J to date. In particular, they took into account both FFA and SSA, and included all relevant energy loss mechanisms for the relativistic electrons. As a result, they were able to adequately reproduce the radio light curves with the standard $s=2$ density profile. In their model, the measured time dependence of the FFA optical depth (i.e. $\delta$) is accounted for by a decrease of $T_{\rm CSM}$ with radius like $T_{\rm CSM} \propto R^{-1}$ (see also Fransson et al. \cite{fra96}). However, as pointed out by Weiler et al. (\cite{wei02}), no evidence for such a radial dependence of $T_{\rm CSM}$ is found in other radio SNe (e.g. SN~1979C and SN~1980K). 

Immler et al. (\cite{imm01}) provided support from X-ray observations to the scenario of a flatter CSM density profile, as they found $s=1.63$ from modeling of the observed X-ray light curve. Their analysis assumes that the X-ray emission arises from the forward, circumstellar shock. But later than $\sim$200~days post-outburst, the X-ray radiation was more likely produced in the SN ejecta heated by the reverse shock (Fransson \& Bj\"ornsson \cite{fra05} and references therein). 

With the model of the present paper, the measured radio light curves can be well explained with the standard $s=2$ assumption, but not with a much flatter CSM density profile. This result is independent of the admittedly uncertain FFA modeling, as we will see in Sect.~4.1 that with $s=1.6$ the {\it optically thin} emission cannot be simultaneously reproduced at all wavelengths in the framework of the model (see Fig.~\ref{fig13}). So the question is how a $\rho_{\rm CSM} \propto R^{-2}$ density profile can be reconciled with the relatively low value of $\delta$ implied by the data. 

A possible explanation is that the absorbing CSM is inhomogeneous. Both radio (Weiler et al. \cite{wei90}) and optical and UV (e.g. Tran et al. \cite{tra97}) observations show that the wind material lost from SN progenitors can be clumpy and/or filamentary. FFA of the radio emission by a nonuniform CSM can be accounted for by an attenuation factor of the form $[1-\exp(-\tau_{\rm CSM}^{\rm clumps})]/\tau_{\rm CSM}^{\rm clumps}$, where $\tau_{\rm CSM}^{\rm clumps}$ is the maximum of the optical depth distribution, which depends on the number density and geometric cross section of the clumps of wind material (Natta \& Panagia \cite{nat84}; Weiler et al. \cite{wei02}). Weiler et al. (\cite{wei07}) performed an overall fit to all of the measured radio light curves of SN~1993J using a parametrized model that takes into account both SSA and FFA, and includes both attenuation by a homogeneous and inhomogeneous CSM (see Appendix~A). In their best-fit model ({\it solid lines} in Fig.~\ref{fig2}), FFA is mainly due to the clumpy CSM, the corresponding optical depth, $\tau_{\rm CSM}^{\rm clumps}$, being much larger than $\tau_{\rm CSM}^{\rm homog}$ in the optically thick phase for all wavelengths (see Table~4 in Weiler et al. \cite{wei07}). 

In view of these results, we are going to neglect in first approximation the attenuation of the radio emission by the homogeneous component of the CSM in front of the attenuation by the clumpy CSM, which will allow us to derive in a simple way and self-consistently the mass loss rate of the progenitor star and the structure of the radio emission region (see below). We anticipate, however, that the adopted simple FFA model will not allow us to accurately reproduce the rising branches of the light curves in the optically thick phase. 

Following Weiler et al. (\cite{wei86,wei02}), the optical depth produced by FFA in a clumpy presupernova wind can be written as 
\begin{equation}
\tau_{\rm CSM}^{\rm clumps} \cong \bigg({\dot{M}_{\rm RSG} \over 4 \pi f_{\rm cl} u_w m_{\rm H}}\bigg)^2 \bigg({1+2X \over 1+4X}\bigg) \bigg({\kappa_{\rm f-f} \over 3 R_0^3}\bigg) \bigg({t \over 1{\rm~day}}\bigg)^{-3m}~,
\label{eq4}
\end{equation}
with, for a uniform $T_{\rm CSM}$, 
\begin{equation}
\kappa_{\rm f-f} = 6.34 \times 10^{-29} \bigg({\nu \over 5{\rm~GHz}}\bigg)^{-2.1} \bigg({T_{\rm CSM} \over 2 \times 10^5{\rm~K}}\bigg)^{-1.35}~~{\rm cm}^5.
\label{eq5}
\end{equation}
Here, $\dot{M}_{\rm RSG}$ and $u_w$ are, respectively, the mass loss rate and wind terminal velocity of the red supergiant progenitor, $m_{\rm H}$ is the mass of a hydrogen atom, $X$ is the He to H abundance ratio in the presupernova wind, and $f_{\rm cl}$ is a factor that depends on the number and geometrical properties of the clumps. Weiler et al. (\cite{wei02}) argue that $f_{\rm cl} \approx 0.67$ (resp. $f_{\rm cl} \sim 0.16$) for attenuation by a statistically large (resp. small) number of clumps along the line of sight. By equating $\tau_{\rm CSM}^{\rm clumps}$ from Eq~(\ref{eq4}) with its expression in the parametric model of Weiler et al. summarized in Appendix~A (Eq.~\ref{eqa6}), one can derive a relation for the progenitor mass loss rate as a function of the normalization parameter $K_3$:
\begin{eqnarray}
\dot{M}_{\rm RSG} & = & 2.2 \times 10^{-7} \sqrt{K_3} \bigg({f_{\rm cl} \over 0.4}\bigg) \bigg({u_w \over 10{\rm~km~s}^{-1}}\bigg) \bigg({T_{\rm CSM} \over 2 \times 10^5{\rm~K}}\bigg)^{0.675} \nonumber \\ 
& \times & \bigg({R_0 \over 3.49 \times 10^{14}{\rm~cm}}\bigg)^{1.5}~~M_\odot~{\rm yr}^{-1},
\label{eq6}
\end{eqnarray}
where we adopted $X=0.3$ (e.g. Shigeyama et al. \cite{shi94}). In the following we use $u_w=10$~km~s$^{-1}$ and $T_{\rm CSM}=2 \times 10^5$~K, the latter value being based on the photoionization calculations performed for SN~1993J by Fransson et al. (\cite{fra96}). We note, however, that according to these calculations $T_{\rm CSM}$ was higher during the first $\sim$10--20~days post-outburst (see Fig.~11 of Fransson et al. \cite{fra96}). 

FFA in the progenitor wind of SN~1993J is estimated below from the following iterative process. First the synchrotron emission as a function of time after outburst is calculated from a set of reasonable initial parameters. External FFA is then estimated from an overall fit of the calculated light curves to the radio data, using Eqs.~(\ref{eqa3}) and (\ref{eqa6}) for the CSM attenuation factor with fixed $\delta'=-3m$ (see Eq.~\ref{eq4}). Thus, the only free parameter is $K_3$. The corresponding progenitor mass loss rate is obtained from Eq.~(\ref{eq6}). The derived wind density just upstream from the forward shock, $\rho_u=\dot{M}_{\rm RSG}/(4 \pi R_s^2 u_w)$, is then used to calculate the shock properties and the associated synchrotron emission. The process is continued until convergence is reached. 

\subsection{Time evolution of the mean magnetic field in the synchrotron-emitting region}

   \begin{figure*}
   \centering
   \includegraphics[width=0.7\textwidth]{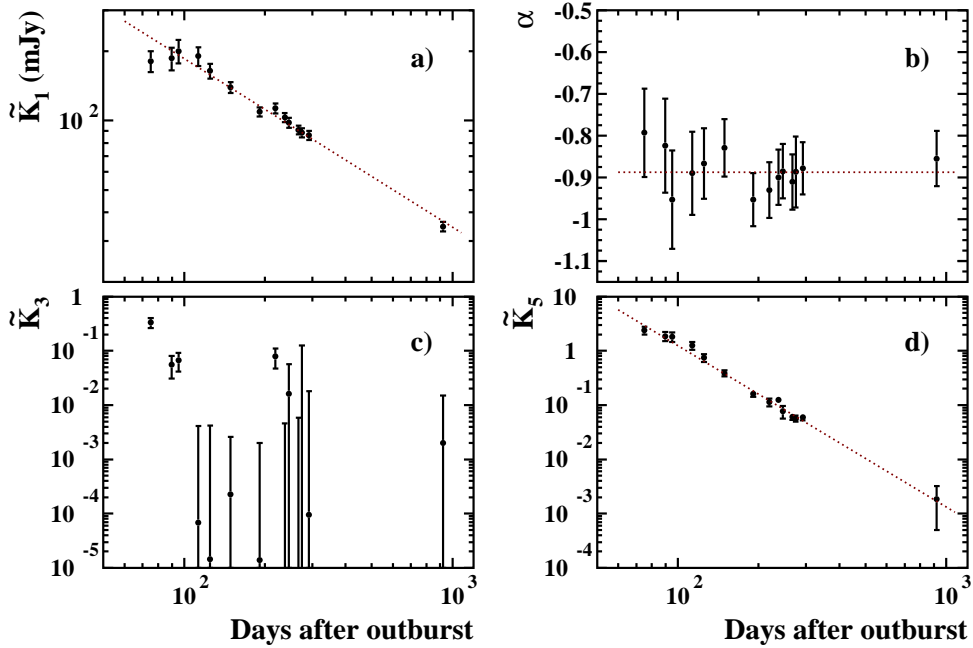}
      \caption{Evolution of the parameters $\tilde{K_1}$, $\alpha$, $\tilde{K_3}$ and $\tilde{K_5}$ obtained from fits to individual VLA spectra (see text). Formally, $\tilde{K_1}$  represents the unabsorbed flux density at $\nu=5$~GHz, $\alpha$ the corresponding synchrotron spectral index, $\tilde{K_3}$ and $\tilde{K_5}$ the FFA and SSA optical depths, respectively, also at $\nu=5$~GHz. The dotted lines show least-squares fit to the extracted parameters (see text).
              }
         \label{fig2p}
   \end{figure*}

In the original model of Chevalier (\cite{che82b}) for the radio emission from SNe, the magnetic energy density in the radio-emitting shell is assumed to scale as the total postshock energy density ($\propto \rho_u V_s^2$), such that the strength of the mean magnetic field $\langle B \rangle \propto t^{-1}$ for $s=2$. Later on, Chevalier (\cite{che96,che98}) also considered that the postshock magnetic field could result from the compression of the circumstellar magnetic field, which would imply $\langle B \rangle \propto R_s^{-1} \propto t^{-m}$. Because SSA plays an important role in the radio emission from SN~1993J, the evolution of the magnetic field can be estimated from the measured light curves. Fransson \& Bj\"ornsson (\cite{fra98}) found, however, the two scaling laws $\langle B \rangle \propto t^{-1}$ and $\langle B \rangle \propto R_s^{-1}$ to be compatible with the data for SN~1993J available at that time. 

We re-estimate here the temporal evolution of $\langle B \rangle$ using the fitting model of Weiler et al. (\cite{wei86,wei02}) together with the simple formalism given in Appendix~A. We analyze individual radio spectra at various dates of observation to determine the time variation of the fitted parameters. The method is similar to the one previously employed by Fransson \& Bj\"ornsson (\cite{fra98}). 

We use observations made with the Very Large Array (VLA) in which a non-zero flux density was measured at at least five wavelengths at the same time (1.2, 2, 3.6, 6, and 20~cm). The corresponding data were obtained later than 75~days  after explosion (see Fig.~\ref{fig2}). We also use the data at 90~cm taken at day 922.7 post-outburst. Each radio spectrum is fitted with the function $F_\nu= \tilde{K_1} (\nu / 5{\rm~GHz})^\alpha A_{\rm CSM}^{\rm clumps} A_{\rm SSA}$, where the attenuation factors are given by Eqs.~(\ref{eqa3}) and (\ref{eqa4}), with the corresponding optical depths $\tau_{\rm CSM}^{\rm clumps} = \tilde{K_3} (\nu / 5{\rm~GHz})^{-2.1}$ and $\tau_{\rm SSA} = \tilde{K_5} (\nu / 5{\rm~GHz})^{\alpha-2.5}$. The fitting function thus contains four free parameters: $\tilde{K_1}$, $\alpha$, $\tilde{K_3}$, and $\tilde{K_5}$. 

The best-fit parameters are shown in Figure~\ref{fig2p}. We took into account only the spectral fits of relatively good quality, with $\chi^2_\nu<2$. It was checked that the final result is not strongly dependent on this data selection. The FFA optical depth $\tilde{K_3}$ was found to be compatible with zero at most epochs. No regular evolution of this parameter can be deduced from the fitting results (Fig.~\ref{fig2p}c). On the other hand, the temporal evolutions of $\tilde{K_1}$ and $\tilde{K_5}$ can be well described by power-law fits (Figs.~\ref{fig2p}a and \ref{fig2p}d). We found
\begin{eqnarray}
\tilde{K_1} & = & (185.3 \pm 5.1) \bigg({t \over 100{\rm~days}}\bigg)^\beta~~{\rm mJy,~with~} \beta=-0.731 \pm 0.026 
\nonumber \\ 
\tilde{K_5} & = & (1.253 \pm 0.090) \bigg({t \over 100{\rm~days}}\bigg)^{\delta''}{\rm ,~with~} \delta''=-2.971 \pm 0.094, 
\nonumber \\
\label{eq7}
\end{eqnarray}
with reduced $\chi^2$ of 1.2 and 1.3, respectively. The spectral index $\alpha$ does not show a significant evolution with time (Fig.~\ref{fig2p}b). Its average constant value is $\alpha=-0.887 \pm 0.020$ ($\chi_\nu^2=0.32$). Then, from Eq.~(\ref{eqa12}) with $\gamma=1-2\alpha=2.774 \pm 0.040$, we get
\begin{equation}
\langle B \rangle = (2.4 \pm 1.0) \bigg({t \over 100{\rm~days}}\bigg)^b {\rm ~G,~with~} b=-1.16 \pm 0.20. 
\label{eq7p}
\end{equation}
We have neglected here the correlations between the various fitted parameters in the error determination, which is a good approximation given that the error in the magnetic field at day 100 mainly arises from the uncertainty in the source distance (9.4\%) and the error in the power-law index $b$ is dominated by the uncertainty in $\delta''$.  

The mean magnetic field given by Eq.~(\ref{eq7p}) is in good agreement with the one obtained by Fransson \& Bj\"ornsson (\cite{fra98}). But in contrast with the conclusions of these authors, the value of $b$ obtained in the present analysis indicates that the scaling law $\langle B \rangle \propto R_s^{-1}$ can be excluded at the 90\% confidence level. The discrepancy is partly due to the different assumptions made for the SN expansion. Indeed, Fransson \& Bj\"ornsson (\cite{fra98}) assumed $m=1$ for $t<100$~days (i.e. $R_s \propto t$) and $m=0.74$ at later epochs. The FFA model is also different in the two analyses. However, because the present result is restricted to relatively late epochs, it is only weakly dependent of the FFA modeling. 

A "pure" SSA model calculated with $\alpha=-0.887$ and the best-fit power laws for $\tilde{K_1}$ and $\tilde{K_5}$ (Eq.~\ref{eq7}) is shown in Figure~\ref{fig2}. We see that this model does not correctly represent the data taken before day 100. This is partly because free-free attenuation of the radio emission in the CSM was not taken into account. But it is also due to the strong radiative losses suffered by the radio-emitting electrons at early epochs (see Fransson \& Bj\"ornsson \cite{fra98} and Sect.~3 below), which are not included in the parametric formalism of Weiler et al. (\cite{wei02} and references therein) used here. As we will see below, one of the main effects of the electron energy losses is to reduce the flux density at short wavelengths during the transition from the optically thick to the optically thin regime. If this effect is not properly taken into account, the relative contributions of FFA and SSA at early epochs post-outburst cannot be reliably estimated. Consequently, the magnetic field determination for these epochs is uncertain. We assume here that the time evolution of $\langle B \rangle$ determined from day 75 after explosion was the same at earlier times. 

We also see in Figure~\ref{fig2} that the decline with time of the optically-thin emission calculated in the SSA model is too slow as compared to the data at 20~cm. A possible explanation is that the energy distribution of the radiating electrons deviates from a power law of constant spectral index, as implicitly assumed in the above modeling (but see below). 

The derived time dependence of $\langle B \rangle$ is consistent with the scaling law originally adopted by Chevalier (\cite{che82b}), $\langle B \rangle \propto t^{-1}$. In Chevalier's model, the scaling law is based on the assumption that the postshock magnetic field is built up by a turbulent amplification powered by the total available postshock energy density. We will see in Sect.~4.3 that the temporal evolution $\langle B \rangle \propto t^{-1}$ is also to be expected if the magnetic field is amplified by Bell (\cite{bel04})'s nonresonant cosmic-ray streaming instability in the shock precursor region and if furthermore the shock is not strongly modified by the back pressure from the energetic ions. We note that the time dependence $\langle B \rangle \propto t^{-1}$ was also recently reported by Soderberg et al. (\cite{sod08}) for the Type Ibc SN~2008D. Therefore, based on both observational evidences and a theoretical basis, I assume in the following that the postshock magnetic field results from an amplification by the cosmic-ray streaming instability operating in the shock precursor and that the field immediately upstream from the subshock is of the form $B_u=B_{u0}(t/1~{\rm day})^{-1}$, where $B_{u0}$ is a free parameter to be determined from fits to the radio data. 

\subsection{Nonlinear particle acceleration}

\begin{figure*}
   \centering
   \includegraphics[width=0.75\textwidth]{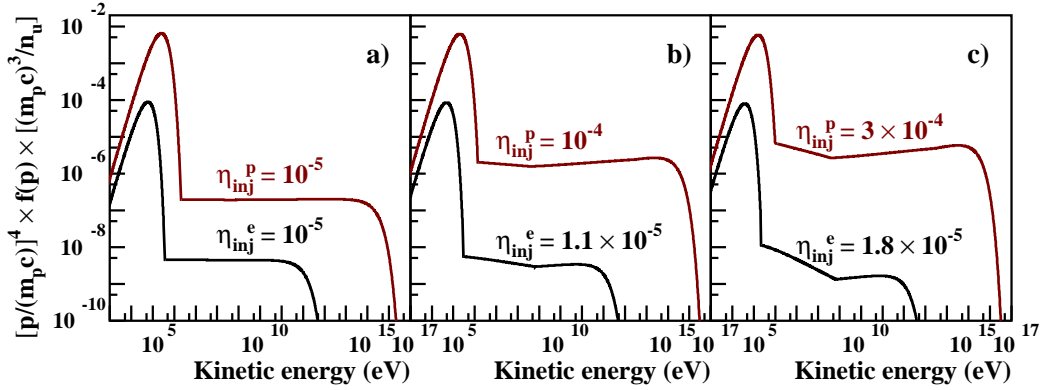}
      \caption{Shocked proton and electron phase-space distributions vs. kinetic energy, at day 1000 after shock breakout. Following Berezhko \& Ellison (\cite{ber99}), the phase-space distribution functions have been multiplied by $[p/(m_pc)]^4$ to flatten the spectra, and by $[(m_pc)^3/n_u]$ to make them dimensionless ($n_u$ is the proton number density ahead of the shock precursor). The upper curves are for protons and the lower ones for electrons. The three sets of injection parameters $\eta_{\rm inj}^p$ and $\eta_{\rm inj}^e$ are those used for the synchrotron calculations shown in Fig.~\ref{fig12}. The magnetic field used for these calculations is $B_u=B_{u0}(t/1~{\rm day})^{-1}=50$~mG (i.e. $B_{u0}=50$~G).
              }
         \label{fig3}
   \end{figure*}

Particle acceleration at the forward shock is calculated with the semianalytic model of nonlinear DSA developed by Berezhko \& Ellison (\cite{ber99}) and Ellison et al. (\cite{ell00}). Although the model strictly applies to plane-parallel, steady state shocks, it has been successfully used by Ellison et al. (\cite{ell00}) for evolving SNRs and more recently by Tatischeff \& Hernanz (\cite{tat07}) to describe the evolution of the blast wave generated in the 2006 outburst of the recurrent nova RS~Ophiuchi. The main feature of this relatively simple model is to approximate the nonthermal part of the shocked proton and electron phase-space distributions as a three-component power law with an exponential cutoff at high momenta, 
\begin{equation}
f_p(p) = a_p p^{-q(p)} \exp(-p/p_{\rm max}^p) 
\label{eq8}
\end{equation}
and
\begin{equation}
f_e(p) = a_e p^{-q(p)} \exp(-p/p_{\rm max}^e) ~,
\label{eq9}
\end{equation}
where the power-law index $q(p)$, which is the same for protons and electrons, can have three different decreasing values in the momentum ranges $p \le m_p c$ ($m_p$ is the proton mass and $c$ the speed of light), $m_p c < p \le 0.01 p_{\rm max}^p$, and $p > 0.01 p_{\rm max}^p$. This piecewise power-law model is intended to account for the upward spectral curvature that results from efficient ion acceleration. The number density of particles per unit energy interval, $N(E)$, is related to the phase-space distribution function by 
$N(E) = 4 \pi p^2 f(p) dp/dE$. 

The maximum proton momentum $p_{\rm max}^p$ is calculated either by time integration of the DSA rate (i.e., shock age limitation) or by equalling the upstream diffusion length to some fraction $f_{\rm esc}$ of the shock radius (i.e., particle escape limitation), whichever produces the lowest value of $p_{\rm max}^p$ (see, e.g., Baring et al. \cite{bar99}). Following, e.g., Ellison \& Cassam-Chena\"i (\cite{ell05}), I take $f_{\rm esc}=0.05$. To estimate the spatial diffusion coefficient, $\kappa$=$\lambda v/3$, the scattering mean free path $\lambda$ of all particles of speed $v$ is assumed to be $\lambda =  \eta_{\rm mfp} r_g$ (Ellison et al. \cite{ell00}), where $r_g$ is the particle gyroradius and $\eta_{\rm mfp}$ is a constant that characterizes the scattering strength. I use $\eta_{\rm mfp}=3$, which is a typical value for young SNRs (Parizot et al. \cite{par06}). The maximum electron momentum $p_{\rm max}^e$ is limited by synchrotron and inverse Compton losses (see Sect.~2.6). 

Given the upstream sonic and Alfv\'en Mach numbers of the shock, which can be readily calculated from $V_s$, $T_{\rm CSM}$, $\rho_u$, and $B_{u0}$ (see below), the proton distribution function (i.e. the normalization $a_p$ and power-law index $q(p)$) is determined by an arbitrary injection parameter $\eta_{\rm inj}^p$, which is the fraction of total shocked protons in protons with momentum $p$$\geq$$p_{\rm inj}^p$ injected from the postshock thermal pool into the DSA process. The work of Blasi et al. (\cite{bla05}) allows us to accurately relate the proton injection momentum $p_{\rm inj}^p$ to $\eta_{\rm inj}^p$. 

The normalization of the electron distribution function is obtained from (Ellison et al. \cite{ell00})
\begin{equation}
a_e =a_p {\eta_{\rm inj}^e \over \eta_{\rm inj}^p} \bigg({m_e \over m_p}\bigg)^{(q_{\rm sub}-3)/2} ~,
\label{eq10}
\end{equation}
where $\eta_{\rm inj}^e$ is the electron injection parameter (i.e. the fraction of shocked electrons with superthermal energies), $m_e$ is the electron mass, and $q_{\rm sub}$ the power-law index $q(p)$ for $p \le m_p c$. Nonthermal particles in this momentum range are accelerated at the gas subshock and we have (Berezhko \& Ellison \cite{ber99})
\begin{equation}
q_{\rm sub} =  {3r_{\rm sub} \over r_{\rm sub} -1}~,
\label{eq11}
\end{equation}
where $r_{\rm sub}$ is the compression ratio of the subshock.  

Alfv\'en wave heating of the shock precursor is taken into account from the simple formalism given in Berezhko \& Ellison (\cite{ber99}). However, a small change to the model of these authors is adopted here: the Alfv\'en waves are assumed to propagate isotropically in the precursor region and not only in the direction opposite to the plasma flow, i.e. Eqs.~(52) and (53) of Berezhko \& Ellison (\cite{ber99}) are not used. This is a reasonable assumption  (see, e.g., Bell \& Lucek \cite{bel01}) given the strong, nonlinear magnetic field amplification required to explain the radio emission from SN~1993J. 

Figure~\ref{fig3} shows calculated shocked proton and electron phase-space distributions for three sets of injection parameters ($\eta_{\rm inj}^p$,$\eta_{\rm inj}^e$) that will be used in Sect.~3.2 to model the radio light curves. The thermal Maxwell-Boltzmann components were calculated using the shocked proton temperature $T_s^p$ determined by the nonlinear DSA model (see, e.g., Ellison et al. \cite{ell00}) and arbitrarily assuming the temperature ratio $T_s^e/T_s^p=0.25$. Noteworthy, the nonthermal electron distribution is independent of $T_s^e/T_s^p$ when $\eta_{\rm inj}^e$ is specified (see Eq.~\ref{eq10}) except for the electron injection momentum  
\begin{equation}
p_{\rm inj}^e=p_{\rm inj}^p \bigg({m_e T_s^e \over m_p T_s^p}\bigg)^{1/2}.
\label{eq12}
\end{equation}
The uncertain temperature ratio has thus practically no influence on the modeled radio emission. 

We see in Figure~\ref{fig3}a that for $\eta_{\rm inj}^p=10^{-5}$ the well-known test-particle result $q(p)=4$ is recovered. But for $\eta_{\rm inj}^p \ge 10^{-4}$ (Figs.~\ref{fig3}b and c) the nonlinear shock modification becomes significant. In particular we see that the nonthermal electron distribution steepens below $\sim 1$~GeV with increasing $\eta_{\rm inj}^p$, as a result of the decrease of $r_{\rm sub}$ (Eq.~\ref{eq11}). This is important because the radio emission from SN~1993J is produced by relativistic electrons of energies $<1$~GeV. 

\begin{figure}
   \centering
   \includegraphics[width=0.45\textwidth]{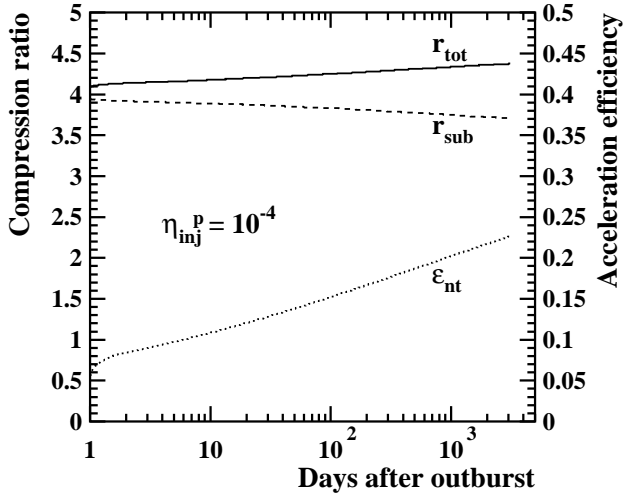}
      \caption{Subshock and total compression ratios at the forward shock ({\it left axis}) and nonthermal energy fraction $\epsilon_{\rm nt}$ ({\it right axis}) as a function of time after outburst, for $\eta_{\rm inj}^p = 10^{-4}$ and the upstream sonic and Alfv\'en Mach numbers of the shock  $M_{S,u}=560 (t/1~{\rm day})^{m-1}$ and $M_{A,u}=9.5$, respectively (see text). 
              }
         \label{fig4}
   \end{figure}

Figure~\ref{fig4} shows calculated subshock and total compression ratios for the case $\eta_{\rm inj}^p = 10^{-4}$ which, as will be shown in Sect.~3.2, provides the best description of the radio light curves. The calculations were performed with the upstream sonic Mach number $M_{S,u} = V_s / c_{S,u} = 560 (t/1~{\rm day})^{m-1}$, given the upstream sound velocity $c_{S,u}=[\gamma_g k T_{\rm CSM}/(\mu m_{\rm H})]^{1/2}=60$~km~s$^{-1}$ for $T_{\rm CSM}=2 \times 10^5$~K. Here $\gamma_g=5/3$ is the adiabatic index for an ideal non-relativistic gas, $k$ is the Boltzmann's constant, and $\mu=(1+4X)/(2+3X)$. Given the magnetic field immediately upstream from the subshock $B_u=B_{u0}(t/1~{\rm day})^{-1}$, the Alfv\'en Mach number $M_{A,u} = V_s / c_{A,u}$ is independent of time; here $c_{A,u}=B_u/\sqrt{4 \pi \rho_u}$ is the Alfv\'en velocity. Anticipating the results presented in Sect.~3, with the best-fit parameter values $B_{u0}=50$~G and $\dot{M}_{\rm RSG}=3.8 \times 10^{-5}~M_\odot$~yr$^{-1}$, we have $M_{A,u}=9.5$. Thus, $M_{A,u} \ll M_{S,u}^2$, which implies that energy should be very efficiently transfered from the accelerated particles to the thermal gas via Alfv\'en wave dissipation in the shock precursor region (Berezhko \& Ellison \cite{ber99}). The resulting increase in the gas pressure ahead of the viscous subshock limits the overall compression ratio, $r_{\rm tot}$, to values close to 4 (i.e. the standard value for a test-particle strong shock). 

However, we see in Figure~\ref{fig4} that the acceleration efficiency $\epsilon_{\rm nt}$ increases with time. This quantity is defined as the fraction of total incoming energy flux, $F_0 \cong 0.5\rho_u V_s^3$, going into shock-accelerated nonthermal particles. At day 3100 after outburst, when the shock has reached the outer boundary of the dense progenitor wind, $\epsilon_{\rm nt}=23$\%. The subshock compression ratio is found to slowly decrease with time, as the shock becomes increasingly modified. Thus, we expect a gradual steepening with time of the electron distribution between $p_{\rm inj}^e$ and $m_pc$ (Eq.~\ref{eq11}). 

Given the high turbulent magnetic field in the shock region, the timescale for diffusive acceleration of electrons in this momentum range is very rapid, $\tau_{\rm acc} \ll 1$~hour. This is much shorter than the characteristic timescale for variation of the shock structure (see Fig.~\ref{fig4}), so it is justified to assume that the spectrum of the radio-emitting electrons at a given time is determined by the instantaneous subshock compression ratio at that time. We note, however, that 
our calculation of the high-energy end of the proton spectrum is not accurate, because the acceleration timescale for the highest energy particles is much longer. 

\subsection{Magnetohydrodynamic evolution of the postshock plasma} 

   \begin{figure}
   \centering
   \includegraphics[width=0.45\textwidth]{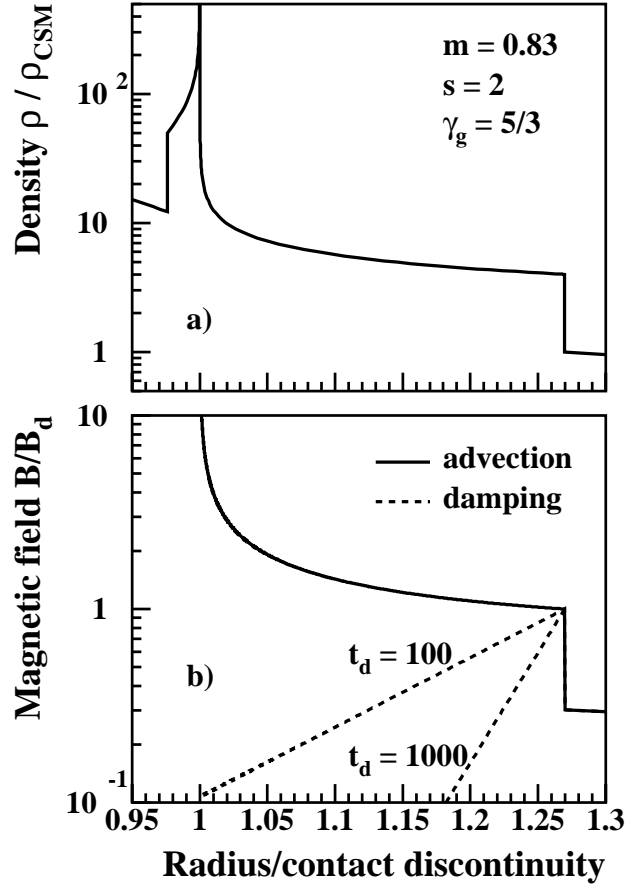}
      \caption{Radial profiles of {\bf (a)} the gas density and {\bf (b)} the  total magnetic field in the shock region. The density is normalized to the upstream value $\rho_u=\dot{M}_{\rm RSG}/(4 \pi R_s^2 u_w)$. The magnetic field is normalized to the immediate postshock value $B_d=B_u[(1+2r_{\rm tot}^2)/3]^{1/2}$. In panel {\bf (b)}, the solid line shows the profile of the postshock magnetic field carried by the flow and the dashed lines that of the damped magnetic field at days 100 and 1000 after outburst (Eqs.~\ref{eq18} and \ref{eq24} with $M_{A,u}=9.5$). The magnetic field is not modeled inside the contact discontinuity.
              }
         \label{fig5}
   \end{figure}

The hydrodynamic evolution of the postshock plasma is calculated using the self-similar model (Chevalier \cite{che82a,che83}; Nadyozhin \cite{nad85}) with the deceleration parameter $m=0.83$ (Sect.~2.1), the standard $s=2$ density profile (Sect.~2.2), and the adiabatic index $\gamma_g=5/3$. Thus, the effects of the back pressure from the accelerated ions on the dynamics of the SNR are neglected. It is a good approximation for SN~1993J given that $4<r_{\rm tot}<4.4$ for the best parameter value $\eta_{\rm inj}^p = 10^{-4}$ (Fig.~\ref{fig4}). The situation is different in older SNRs, such as the remnant of Kepler's (Decourchelle et al. \cite{dec00}) and Tycho's (Warren et al. \cite{war05}) SNe. In these objects, the backreaction of shock-accelerated cosmic rays has more influence on the shock structure, mainly because the magnetic field in the precursor region is much lower than for SN~1993J, such that Alfv\'en wave heating is less important. 

Figure~\ref{fig5}a shows the density profile in the region of interaction between the SN ejecta and the progenitor wind. The locations of the forward and reverse shocks are clearly visible, at 1.27 and 0.976 times the radius of the contact discontinuity, respectively.

The postshock magnetic field is thought to result from the shock compression of the turbulent magnetic field immediately upstream from the subshock, that has been presumably amplified in the precursor region by both resonant (Bell \& Lucek \cite{bel01}) and nonresonant (Bell \cite{bel04}) cosmic-ray streaming instabilities. Following the work of Cassam-Chena\"i et al. (\cite{cas07}) for Tycho's SNR, I make two different assumptions about the postshock magnetic field evolution: one where the turbulent magnetic field is simply carried by the downstream plasma flow (i.e. advected) and another where the magnetic turbulence is rapidly damped behind the blast wave. 

\subsubsection{Advected downstream magnetic field}

The equations describing the evolution of the postshock magnetic field in the case of pure advection in the downstream plasma are given in Cassam-Chena\"i et al. (\cite{cas05}) and references therein. They are reproduced here for sake of convenience. Let the radial and tangential components of the immediate postshock magnetic field be $B_{d,r}$ and $B_{d,t}$, respectively. Assuming that the upstream magnetic field is fully turbulent and isotropic, we have
\begin{eqnarray}
B_{d,r} & = & 1/\sqrt{3} B_u \\ \label{eq13}
B_{d,t} & = & \sqrt{2/3} r_{\rm tot} B_u~. \label{eq14}
\end{eqnarray}
At time $t$ after outburst, the radial and tangential components of the magnetic field in a fluid element with density $\rho(R,t)$ at the downstream position $R$ are given by 
\begin{eqnarray}
B_{r}(R,t) & = & B_{d,r}(t_i) \bigg({R(t) \over R_s(t_i)}\bigg)^{-2} \\ \label{eq15}
B_{t}(R,t) & = & B_{d,t}(t_i) {\rho(R,t) \over \rho_d(t_i)} {R(t) \over R_s(t_i)}~,
\label{eq16}
\end{eqnarray}
where $t_i$ is the earlier time when this fluid element was shocked and $\rho_d(t_i)$ is the immediate postshock density at that time. The total magnetic field is then simply 
\begin{equation}
B(R,t)=\big( B_{r}(R,t)^2 + B_{t}(R,t)^2 \big)^{1/2}. 
\label{eq17}
\end{equation}

The radial profile of the total advected magnetic field is shown in Figure~\ref{fig5}b. Under the assumption of self-similarity, the plotted ratio 
$B(R)/B_d$ is independent of time. 

\subsubsection{Damped downstream magnetic field}

Pohl et al. (\cite{poh05}) suggested that the nonthermal X-ray filaments observed in Galactic SNRs could be localized enhancements of the magnetic field in the blast wave region. In this scenario, the turbulent magnetic field amplified in the shock presursor is thought to be rapidly damped behind the shock front by cascading of wave energy to very small scales where it is ultimately dissipated. 

Assuming a Kolmogorov-type energy cascade of incompressible magnetohydrodynamic (MHD) turbulence, the characteristic damping length can be estimated to be (Pohl et al. \cite{poh05}; see also Cassam-Chena\"i et al. \cite{cas07}) 
\begin{equation}
l_{\rm damp} \approx {5 \over \pi} {u_d \over c_{A,d}}\lambda_{\rm max} \approx 
{5 \over 0.83 \pi} {M_{A,u} \over r_{\rm tot}^{3/2}}\lambda_{\rm max} ~,
\label{eq18}
\end{equation}
where $u_d=V_s / r_{\rm tot}$ is the downstream flow speed in the shock rest frame, $c_{A,d} \cong 0.83 r_{\rm tot}^{1/2} c_{A,u}$ is the immediate postshock Alfv\'en speed, and $\lambda_{\rm max}$ is the maximum wavelength of the magnetic turbulence, which is expected to be of the order of the gyroradius of the maximum energy protons. In a size-limited shock, where the proton maximum energy is determined by upstream particle escape, we have 
\begin{equation}
\lambda_{\rm max} \sim r_{g, \rm max} = { 3 f_{\rm esc} R_s V_s \over \eta_{\rm mfp} c}~,
\label{eq19}
\end{equation}
such that
\begin{equation}
{l_{\rm damp} \over R_s} \sim {5.8 M_{A,u} f_{\rm esc} V_s \over r_{\rm tot}^{3/2} \eta_{\rm mfp} c}~.
\label{eq20}
\end{equation}
For $V_s=3.35 \times 10^4 (t/1~{\rm day})^{m-1}$~km~s$^{-1}$ (Eq.~\ref{eq2}), $f_{\rm esc}=0.05$, $\eta_{\rm mfp}=3$, and $r_{\rm tot} \approx 4$ (Sect.~2.4), one gets
\begin{equation}
{l_{\rm damp} \over R_s} \sim 1.3 \times 10^{-3} M_{A,u} \bigg({t \over 1{\rm~day}}\bigg)^{m-1}~.
\label{eq21}
\end{equation}
We see that for $M_{A,u} \lsim 100$, such a magnetic field damping is expected  
to produce thin radio filaments, which were not observed in SN~1993J (e.g., Bietenholz et al. \cite{bie03}). Therefore, I do not use this model of turbulence damping.

Pohl et al. (\cite{poh05}) also considered that downstream magnetic field damping can result from cascading of fast-mode and Alv\'en waves in background MHD turbulence. In these cases, we have (see also Cassam-Chena\"i et al. \cite{cas07})
\begin{equation}
l_{\rm damp} \sim {1 \over 2 \sqrt{2\pi}} {u_d \over c_{A,d}} \sqrt{\lambda_{\rm max} L}~,
\label{eq22}
\end{equation}
where $L$ is the outer scale of the pre-existing MHD turbulence. Assuming that this quantity is of the order of the diameter of the dense wind bubble blown the red supergiant progenitor of the SN, $L\sim 2 R_{\rm out}=6\times10^{17}$~cm, we obtain
\begin{equation}
{l_{\rm damp} \over R_s} \sim 0.1 M_{A,u} \bigg({t \over 1{\rm~day}}\bigg)^{-1/2}~.
\label{eq23}
\end{equation}
This damping length is thus always larger (for $t<3100$~days) than the one estimated from the general model of Kolmogorov-type energy cascade (Eq.~\ref{eq21}). In the case of damping by cascading of fast-mode and Alfv\'en waves the spatial relaxation of the postshock magnetic field follows an exponential decay law
\begin{equation}
B(R)=B_d \exp\bigg(-{R_s - R \over l_{\rm damp}}\bigg)~.
\label{eq24}
\end{equation}
Postshock magnetic field profiles obtained from Eqs.~(\ref{eq23}) and (\ref{eq24}) with $M_{A,u}=9.5$ are shown in Figure~\ref{fig5}b. 

\subsection{Nonthermal electron energy losses}

The cooling processes that can affect the energy distribution of the nonthermal electrons during the SN expansion were studied by Fransson \& Bj\"ornsson (\cite{fra98}). They found synchrotron cooling to be predominant for electrons radiating at short wavelengths for most of the time, Coulomb cooling to be potentially important at early epochs post-outburst, adiabatic cooling to be dominant for electrons radiating at 20~cm at late epochs and inverse Compton losses due to electron scattering off photons from the SN photosphere to be less important. These results depend, however, on several model parameters, e.g. the progenitor mass loss rate $\dot{M}_{\rm RSG}$ and the magnetic field $B_{u0}$. 

With the best parameter values of the present model, Coulomb cooling is found to be more important than synchrotron cooling only at very early epochs, when the radio emission is still optically thick to SSA (Sect.~4.2). At that time, the energy losses of the shock-accelerated electrons have little if no effect on the radio emission, such that Coulomb cooling can be safely neglected. This contrasts with the model of Fransson \& Bj\"ornsson (\cite{fra98}). 

I use the work of Reynolds (\cite{rey98}; see also Cassam-Chena\"i et al. \cite{cas07}, Appendix B) to calculate the downstream evolution of the electron energy distribution due to synchrotron, inverse Compton, and adiabatic losses. The number density of nonthermal electrons per unit energy interval at kinetic energy $E$ in a fluid element being at the downstream position $R$ at time $t$ can be written as
\begin{equation}
N_e(E,R,t)=N_e(E_i,R_s(t_i),t_i) \bigg({\rho(R,t) \over \rho_d(t_i)}\bigg)^{4/3} \bigg({E_i \over E}\bigg)^2~,
\label{eq25}
\end{equation}
where $N_e(E_i,R_s(t_i),t_i)$ is the electron distribution function at the initial energy $E_i$ produced at the shock at the earlier time $t_i$ when the fluid element was shocked\footnote{We note that in Reynolds (\cite{rey98}) a factor $\alpha_l=\rho(R,t) / \rho_d(t_i)$ is missing in the unnumbered equation just before eq.~(24) and in eq.~(25).}. The change in energy during the expansion is given by 
\begin{equation}
E = \alpha_l^{1/3} { E_i \over 1 + \Theta E_i}~.
\label{eq26}
\end{equation}
Here, $\Theta$ is a radiative loss term that includes both synchrotron and inverse Compton cooling:
\begin{equation}
\Theta = {4 \over 3} {\sigma_{\rm T} c \over (m_ec^2)^2} \int_{t_i}^t \bigg[ {B^2(\tau) \over 8\pi} + U_{\rm rad}(\tau) \bigg] \alpha_l^{1/3}(\tau) d\tau~,
\label{eq27}
\end{equation}
where $\sigma_{\rm T}$ is the Thomson cross section, $B(\tau)$ is the time-dependent magnetic field in the fluid element, and 
\begin{equation}
U_{\rm rad}(\tau) \approx {L_{\rm bol}(\tau) \over 4 \pi c R^2(\tau)}
\label{eq28}
\end{equation} 
is the energy density in the radiation field, $L_{\rm bol}$ being the bolometric luminosity. 

   \begin{figure}
   \centering
   \includegraphics[width=0.5\textwidth]{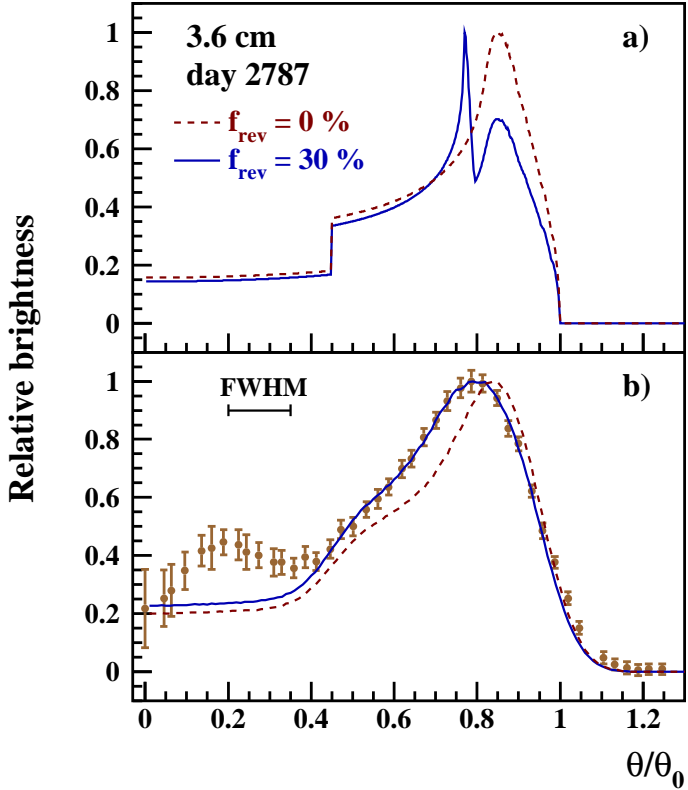}
      \caption{Relative brightness of the 3.6~cm emission from SN~1993J at day 2787 after outburst, as a function of angular radius. In the calculations, the outer angular radius $\theta_0$ is set equal to $R_s/D$. Panel~{\bf (a)} shows two unconvolved profiles calculated for $f_{\rm rev}=0$ (i.e. no contribution from electrons accelerated at the reverse shock) and $f_{\rm rev}=30$\%, assuming in both cases that the magnetic field is advected behind the blast wave, with $B_{u0}=50$~G ($B_u=18$~mG at day 2787) and $\rho_{\rm abs}=0.45R_s$. Panel~{\bf (b)} shows the two profiles convolved with the instrumental resolution of 0.70~mas FWHM (0.15~$\theta_0$ at day 2787) compared to the VLBI data of Bietenholz et al. (\cite{bie03}). 
              }
         \label{fig6}
   \end{figure}

Fransson \& Bj\"ornsson (\cite{fra98}) showed that during the first $\sim$100~days after outburst, the radiation energy density at the forward shock position was dominated by emission from the SN ejecta. They approximated the bolometric luminosity of SN~1993J at these early epochs by
\begin{equation}
L_{\rm bol} \approx 4 \times 10^{42} \bigg({t \over 10~{\rm days}}\bigg)^{-0.9}~~{\rm erg~s}^{-1}.
\label{eq29}
\end{equation}
The ratio of the energy densities of magnetic to seed photon fields in the immediate postshock region is then
\begin{equation}
{U_B \over U_{\rm rad}}={B_d^2 \over 8\pi U_{\rm rad}} \approx \bigg({B_{u0} \over 40~{\rm G}}\bigg)^2 \bigg({t \over 1{\rm~day}}\bigg)^{0.56}.
\label{eq30}
\end{equation}
This result shows that Compton cooling can play a role soon after outburst depending on the magnetic field strength, but  that synchrotron cooling is expected to become more important after some time. A similar result was found by Chevalier \& Fransson (\cite{che06}) for Type~Ib/c SNe. 

\subsection{Radio synchrotron emission}

Once the electron energy distribution and the magnetic field as a function of downstream position are determined as explained above, the synchrotron brightness profile and the total flux density at a given time can be calculated as described in Appendix~B. 

   \begin{figure}
   \centering
   \includegraphics[width=0.5\textwidth]{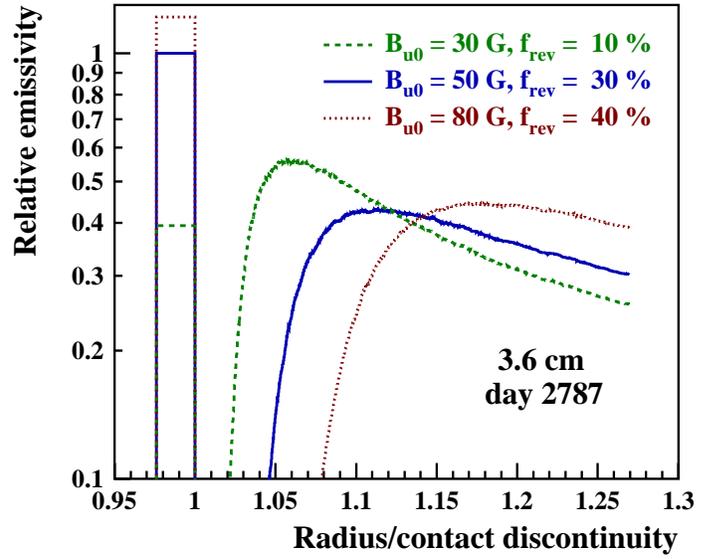}
      \caption{Calculated radial profiles of the relative emissivity at 3.6~cm and $(t/1~{\rm day})=2787$, for $B_{u0}=30$~G and $f_{\rm rev}=10$\% ({\it green dashed lines}), $B_{u0}=50$~G and $f_{\rm rev}=30$\% ({\it blue solid lines}), and $B_{u0}=80$~G and $f_{\rm rev}=40$\% ({\it red dotted lines}). The shell of uniform emissivity near the contact discontinuity is intended to account for the emission produced by electrons accelerated at the reverse shock (see text).
              }
         \label{fig7}
   \end{figure}

\section{Results}

For a given assumption about the evolution of the postshock magnetic turbulence (advection or damping), the model has four main parameters to be fitted to the radio data: the mass loss rate of the red supergiant progenitor, $\dot{M}_{\rm RSG}$, the normalization of the upstream magnetic field, $B_{u0}$, and the proton and electron injection parameters, $\eta_{\rm inj}^p$ and $\eta_{\rm inj}^e$. 
In addition, the radius of the absorbing disk, $\rho_{\rm abs}$, that accounts for FFA of the radio emission by the inner SN ejecta (see Appendix~B) mainly influences the brightness profile as a function of angular radius. We first compare calculated synchrotron profiles with the high-resolution profile at 3.6~cm measured by Bietenholz et al. (\cite{bie03}) from VLBI observations. We then study the radio light curves reported by Weiler et al. (\cite{wei07}). 

\subsection{The radial brightness profile}

To study the radial brightness profile of SN~1993J with the highest angular resolution, Bietenholz et al. (\cite{bie03}) produced a composite image at 8.4~GHz from  VLBI observations performed at $t=2080$, 2525, and 2787~days after explosion. The data were appropriately scaled to take into account the SN expansion and then averaged. The resulting brightness profile versus angular radius is shown in Figure~\ref{fig6} together with two calculated profiles. Here and in the following, we set the radius of the inner opaque disk accounting for FFA of the radio emission from the side of the shell moving away from us (Appendix~B) to $\rho_{\rm abs}=0.45R_s$. We see in Figure~\ref{fig6} that this simple model of absorption by the SN ejecta is consistent with the data for $\theta/\theta_0>0.4$, but underestimates the observed emission for $\theta/\theta_0<0.4$. The difference might arise from incomplete absorption of the radio waves in the inner ejecta, possibly because the latter are filamentary (see Bietenholz et al. \cite{bie03}). However, the excess of emission at $\theta/\theta_0 \sim 0.2$ accounts for only $\sim$3\% of the total flux density and it has been neglected so as to limit the number of free parameters. 

   \begin{figure}
   \centering
   \includegraphics[width=0.5\textwidth]{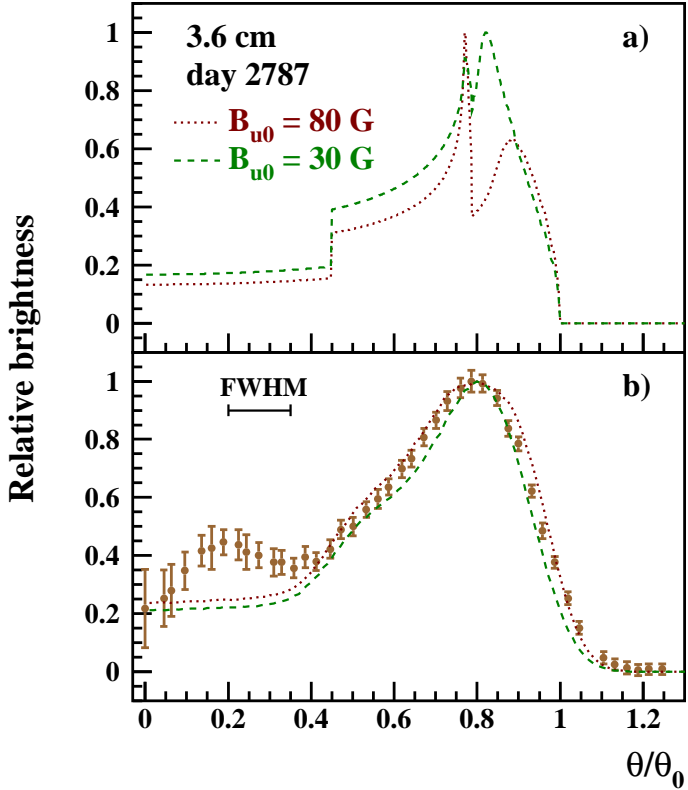}
      \caption{Same as Fig.~7 but for $B_{u0}=30$~G, $f_{\rm rev}=10$\% ({\it green dashed lines}) and $B_{u0}=80$~G, $f_{\rm rev}=40$\% ({\it red dotted lines}).}
         \label{fig8}
   \end{figure}

   \begin{figure}
   \centering
   \includegraphics[width=0.5\textwidth]{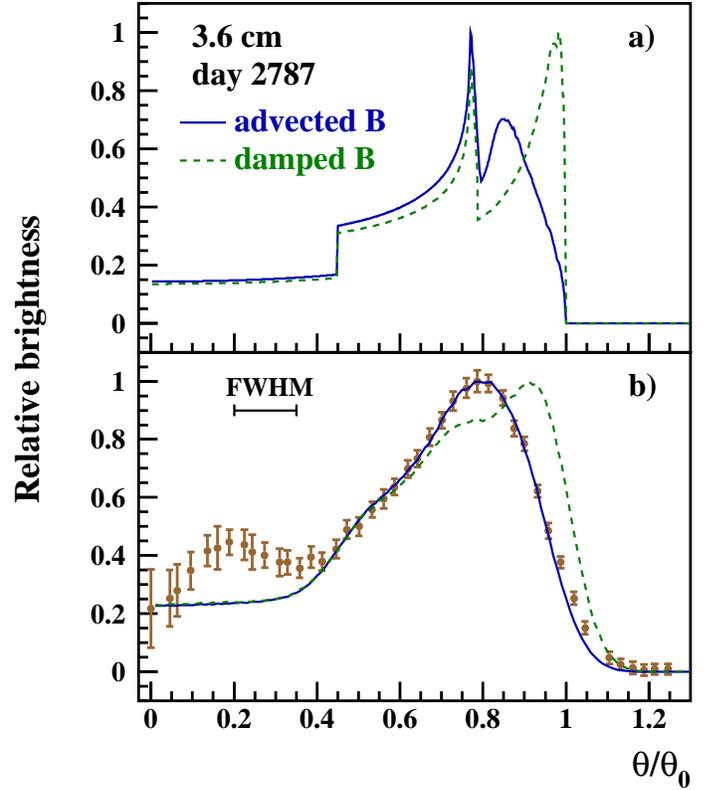}
      \caption{Same as Fig.~7 but for the damping of the magnetic turbulence downstream ({\it green dashed lines}) compared to the advection of the shocked plasma ({\it blue solid lines}). In both cases, $B_{u0}=50$~G and $f_{\rm rev}=30$\%.}
         \label{fig9}
   \end{figure}

The radial brightness profile is mainly sensitive to the strength and profile of the magnetic field behind the blast wave\footnote{In contrast, the radio morphology of older SNRs, such as Tycho's remnant, also depends on the proton injection parameter, $\eta_{\rm inj}^p$, because in these objects an increase in the injection efficiency produces a narrower interaction region (Cassam-Chena\"i et al. \cite{cas07}). This is not the case in SN~1993J because of the much higher magnetic field strength in the shock precursor region, which limits the backreaction of the shock-accelerated  nuclei on the shock structure (see Sect.~2.4).}. The two calculated profiles shown in Figure~\ref{fig6} are for $B_{u0}=50$~G and the case of pure advection of the field in the downstream plasma. We see in Figure~\ref{fig6}b that the model with only electrons accelerated at the forward shock ({\it red dashed curve}) produces a peak at $\theta/\theta_0 \sim 0.85$ that is both slightly shifted and too narrow as compared to the observed one. There is clearly a deficit of emission at $\theta/\theta_0 < 0.8$, i.e. from a region close to the contact discontinuity that marks the border between shocked ejecta and shocked CSM (see Fig.~\ref{fig5}). 

As shown in Figure~\ref{fig7}, the radial emissivity of electrons accelerated at the blast wave is cut off before it reaches the contact discontinuity. This is due to the strong radiative losses suffered by the electrons that have been accelerated at the earliest epochs. The cutoff exists whatever the magnetic fied strength $B_{u0}$; it is mainly produced by Compton cooling for $B_{u0} \lsim 10$~G (see Eq.~\ref{eq30}). Thus, the missing component of synchrotron emission is probably coming from another source of accelerated electrons, most likely the reverse shock. The efficiency of particle acceleration at reverse shocks in SNRs is poorly known, because it depends on the unknown amplification of the magnetic field in the unshocked ejecta material (Ellison et al. \cite{ell05b}). However, recent observations of SNRs provide clear evidence for synchrotron emission associated with reverse shocks (see Helder \& Vink \cite{hel08} for Cassiopeia~A). Here, we simply model this emission by a shell of uniform emissivity between the reverse shock and the contact discontinuity (Fig.~\ref{fig7}), whose normalization is fitted to the radial brightness profile measured by Bietenholz et al. (\cite{bie03}). We use as normalization factor the integral of the radial emissivity relative to that for the forward shock:
\begin{equation}
f_{\rm rev}={(R_{\rm CD}-R_{\rm RS})\epsilon_\nu^{\rm RS} \over \int_{R_{\rm CD}}^{R_{\rm FS}} \epsilon_\nu^{\rm FS}(R)dR}~, 
\label{eq31}
\end{equation}
where $R_{\rm CD}$, $R_{\rm RS}$, and $R_{\rm FS}$ are the radii of the contact discontinuity, reverse and forward shocks, respectively. We see in Figure~\ref{fig6} that for $\theta/\theta_0>0.4$ an excellent fit to the measured brightness profile is obtained for $B_{u0}=50$~G and $f_{\rm rev}=30$\%. After integration along the line-of-sight, the synchrotron emission from electrons accelerated at the reverse shock is found to contribute for 17\% of the total detected flux density. 

   \begin{figure*}
   \centering
   \includegraphics[width=0.65\textwidth]{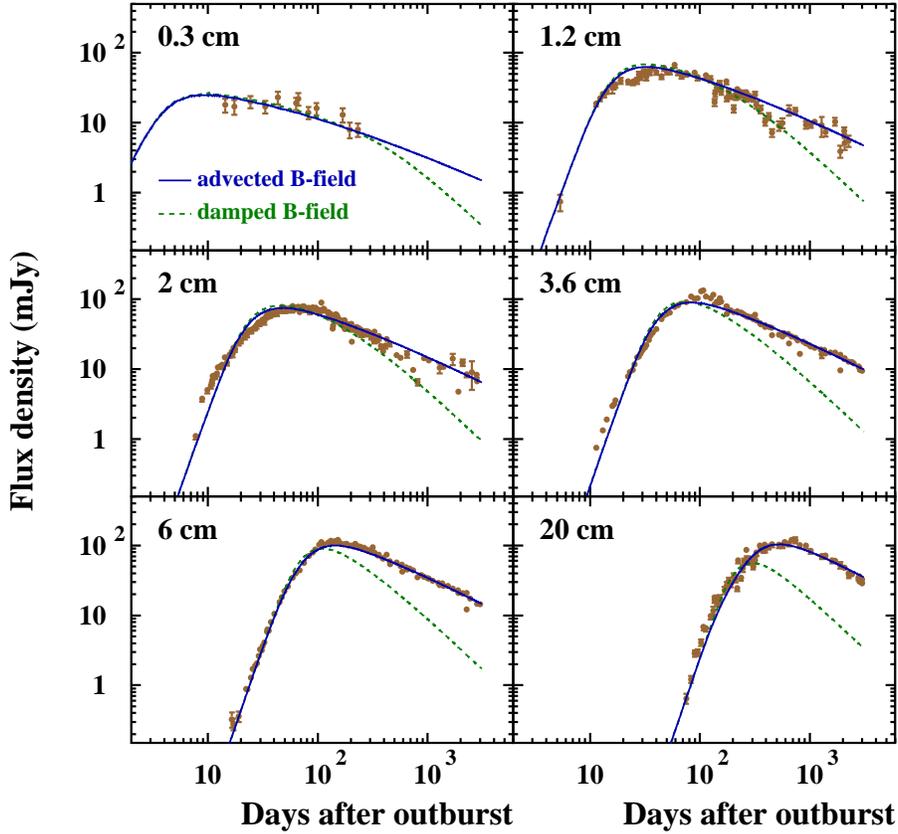}
      \caption{Radio light curves for SN~1993J at 0.3, 1.2, 2, 3.6, 6, and 20~cm. The data are from Weiler et al. (\cite{wei07}) and references therein. The blue solid lines represent the best-fit model, which is obtained for $\dot{M}_{\rm RSG}=3.8 \times 10^{-5}~M_\odot$~yr$^{-1}$, $B_{u0}=50$~G, $\eta_{\rm inj}^p=10^{-4}$, $\eta_{\rm inj}^e=1.1\times10^{-5}$, and pure advection of the postshock magnetic field. The green dashed lines show the case of magnetic turbulence damping in the downstream plasma. 
              }
         \label{fig10}
   \end{figure*}

Figure~\ref{fig7} also shows that the radial emissivity profile decreases more rapidly behind the blast wave as $B_{u0}$ increases, which is due to the synchrotron energy losses. Brightness profiles calculated for $B_{u0}=30$ and 80~G are compared to the data in Figure~\ref{fig8}. We adjusted the value of $f_{\rm rev}$ so as to match the maximum of the relative brightness to the observed one at $\theta/\theta_0 = 0.8$. But we see that the width of the broad emission peak is not well fitted: the calculated emission profile is too narrow (resp. too broad) for the low (resp. high) value of $B_{u0}$. Thus, the high-resolution brightness profile of Bietenholz et al. provides a first indication of the magnetic field strength in the blast wave region: $11<B_u<29$~mG at day 2787 post-outburst ($30<B_{u0}<80$~G). 

The measured profile also constraints the evolution of the postshock magnetic field. Figure~\ref{fig9} shows that, when the magnetic field is damped behind the blast wave, the brightness angular distribution is shifted toward the outer edge of the radio shell and the resulting profile is clearly not consistent with the data. This conclusion is independent of both the value of $B_u$ and the reverse shock contribution. In particular, the brightness profile becomes clearly too broad when $f_{\rm rev}$ is increased so as to set the maximum of the emission at $\theta/\theta_0 = 0.8$. 

The contact interface between the SN ejecta and the shocked CSM is thought to be Rayleigh-Taylor unstable and it has been suggested that the associated turbulence can amplify the postshock magnetic field (Chevalier et al. \cite{che92}; Jun \& Norman \cite{jun96}). This would increase the synchrotron losses near the contact discontinuity. Consequently, the position of the cutoff in the radial emissivity would likely be shifted to larger radii with respect to the case with no amplification of the postshock magnetic field (Fig.~\ref{fig7}). As a result, a larger contribution of synchrotron radiation from electrons accelerated at the reverse shock would probably be needed to reproduce the measured brightness profile. 

However, the shocked CSM is expected to be strongly magnetized due to the field amplification by the cosmic-ray streaming instability operating in the forward shock precursor and the postshock magnetic field profile is expected to be dominated by the tangential component (see Eqs.~18 and \ref{eq16}). Such a strong magnetic field could decrease the growth of the Rayleigh-Taylor instability and limit a possible additional amplification of the field by the turbulence associated with this instability (Jun et al. \cite{jun95}). This aspect of the magnetohydrodynamic evolution of the postshock plasma certainly deserves further studies. Here, we assume that the postshock magnetic field is not significantly amplified by the Rayleigh-Taylor instability. 

\subsection{Radio light curves}

In the modeling of the radio light curves, we take into account the emission from electrons accelerated at the reverse shock assuming $f_{\rm rev}=30$\% at all times. As discussed above, this emission is then estimated to contribute to a maximum of 17\% of the total flux density. This number is an upper limit, because the synchrotron radiation from the inner shock is strongly attenuated when the emission associated with the forward shock is optically thick to SSA. We also assume $\rho_{\rm abs}=0.45R_s$ at all epochs. The resulting uncertainty on the flux density is negligible. 

   \begin{figure*}
   \centering
   \includegraphics[width=0.65\textwidth]{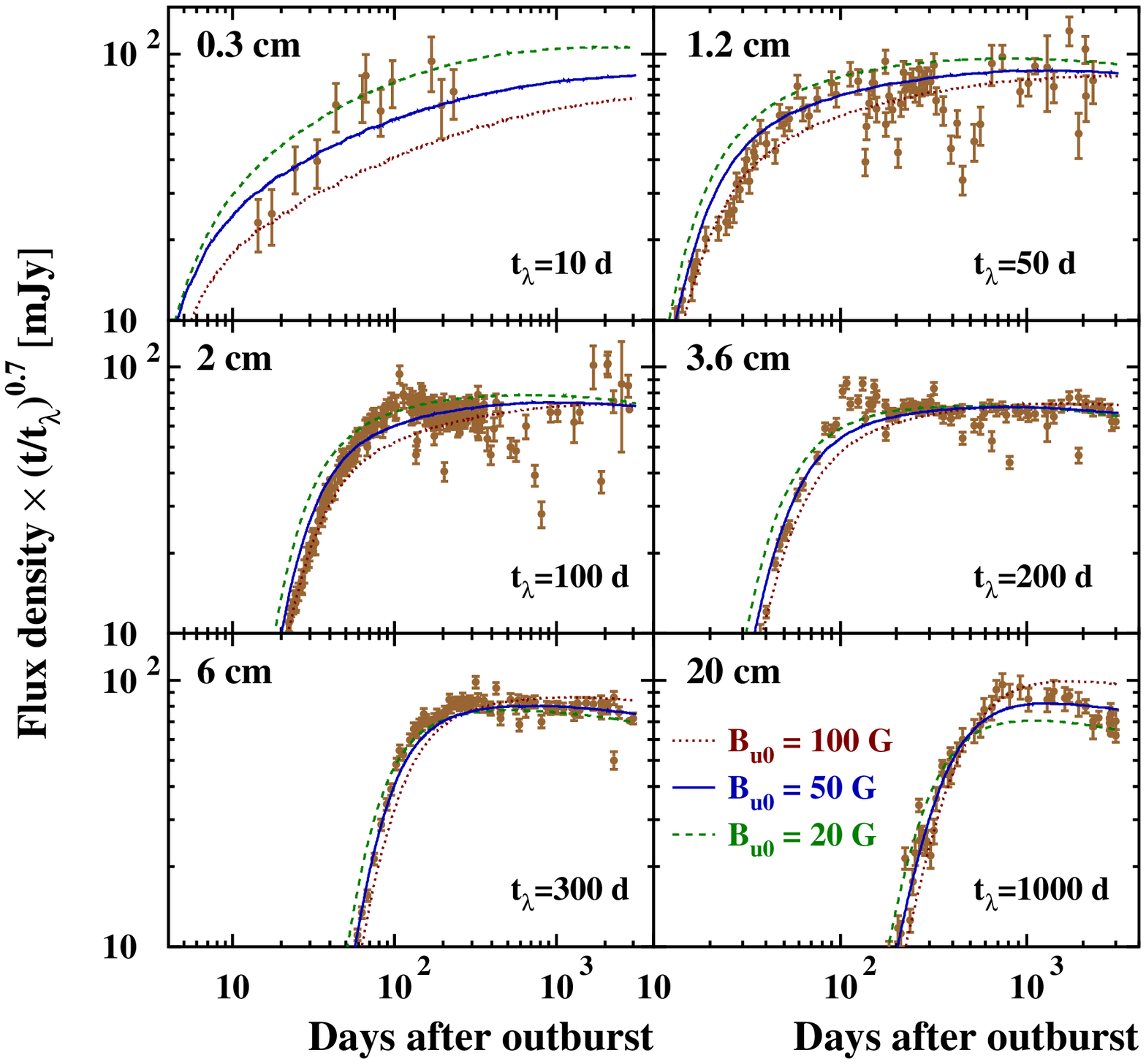}
      \caption{Radio light curves of SN~1993J with the flux density multiplied by $(t/t_\lambda)^{0.7}$. The values of $t_\lambda$ for each wavelength are indicated in the Figure. {\it Green dashed lines}: $B_{u0}=20$~G and $\eta_{\rm inj}^e=2.3\times10^{-5}$; {\it blue solid lines}: $B_{u0}=50$~G and $\eta_{\rm inj}^e=1.1\times10^{-5}$; {\it red dotted lines}: $B_{u0}=100$~G and $\eta_{\rm inj}^e=8\times10^{-6}$. The other parameters are $\dot{M}_{\rm RSG}=3.8 \times 10^{-5}~M_\odot$~yr$^{-1}$ and $\eta_{\rm inj}^p=10^{-4}$. 
              }
         \label{fig11}
   \end{figure*}

   \begin{figure*}
   \centering
   \includegraphics[width=0.65\textwidth]{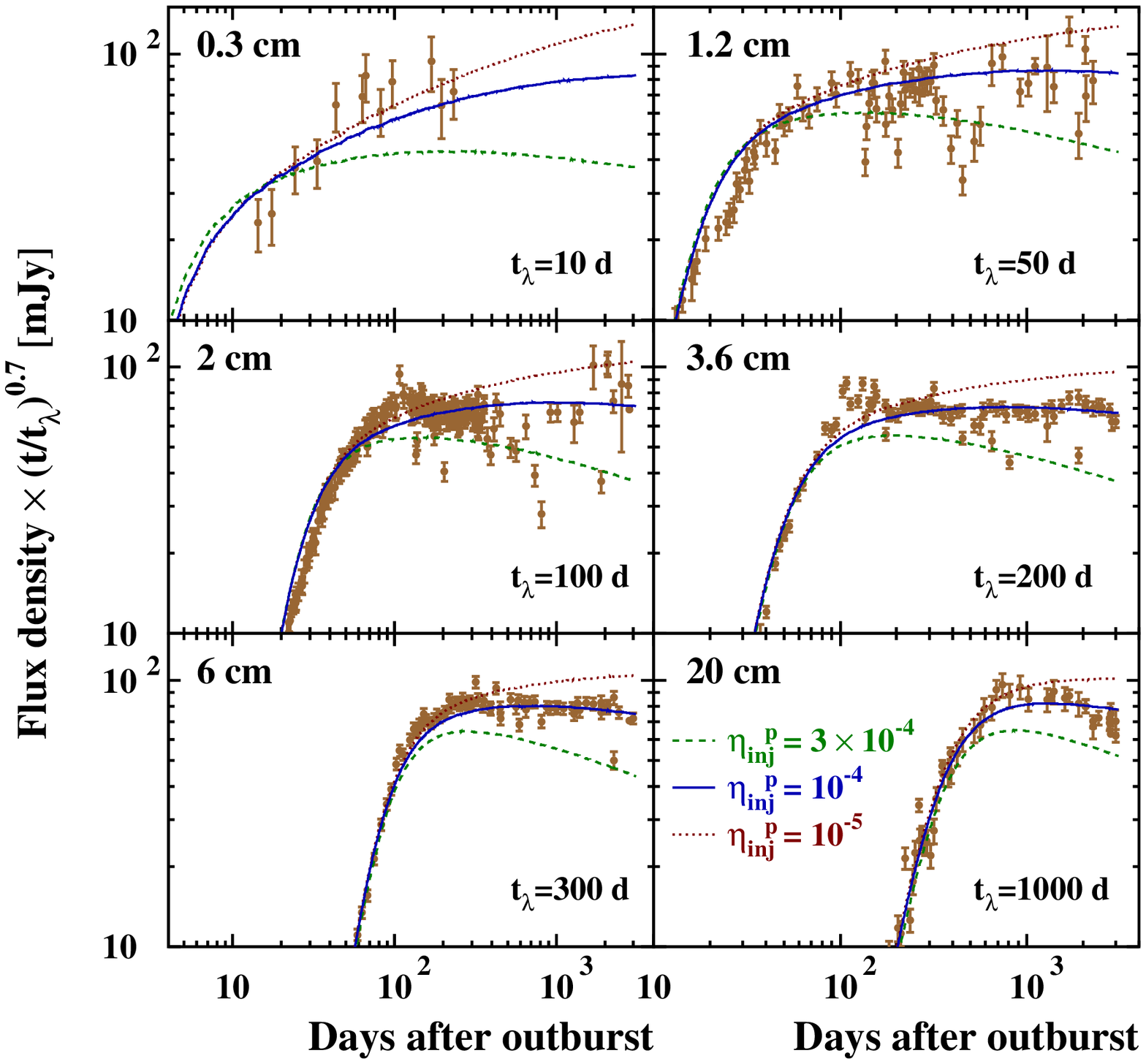}
      \caption{Same as Fig.~12 but for various values of the injection parameters: $\eta_{\rm inj}^p=\eta_{\rm inj}^e=10^{-5}$ ({\it red dotted lines}); $\eta_{\rm inj}^p=10^{-4}$ and $\eta_{\rm inj}^e=1.1\times10^{-5}$ ({\it blue solid lines}), and $\eta_{\rm inj}^p=3\times10^{-4}$ and $\eta_{\rm inj}^e=1.8\times10^{-5}$ ({\it green dashed lines}). The other parameters are $\dot{M}_{\rm RSG}=3.8 \times 10^{-5}~M_\odot$~yr$^{-1}$ and $B_{u0}=50$~G. 
              }
         \label{fig12}
   \end{figure*}

Light curves calculated with the best-fit model are shown in Figure~\ref{fig10}. They were obtained for $\dot{M}_{\rm RSG}=3.8 \times 10^{-5}~M_\odot$~yr$^{-1}$, $B_{u0}=50$~G, $\eta_{\rm inj}^p=10^{-4}$, $\eta_{\rm inj}^e=1.1\times10^{-5}$, and pure advection of the postshock magnetic field. We see that the general features of the data are fairly well represented. But the associated reduced $\chi^2$ is $\chi^2_\nu=7.9$ ($\chi^2=4330$ for 554 degrees of freedom), which is comparable to the one of the best-fit semi-empirical model of Weiler et al. (\cite{wei07}): $\chi^2_\nu=8.1$\footnote{These high values of $\chi^2_\nu$ are partly due to a few data points that strongly deviate from neighbouring points and appear to be incompatible with the general trend observed at each wavelength.}. However, in the comparison of the $\chi^2$ values, one could note that the formalism of Weiler et al. has nine free parameters (Appendix~A), against only four in the present model. Interestingly, the present model is also good at representing the 12 data points at 0.3~cm ($\chi^2=15.5$), which is not the case for the parametric model of Weiler et al. ($\chi^2=159$; see Fig.~\ref{fig2}). The 0.3~cm flux density between $\sim$10 and $\sim$40~days after outburst is lower in the present calculations than in previous ones, because the strong synchrotron losses suffered by the radiating electrons during the early SN expansion are now taken into account. The difference is more pronounced at the shortest wavelength, because this emission is produced by higher-energy electrons (on average), for which the synchrotron cooling time is lower ($\tau_{\rm syn} \propto 1/\Gamma$, $\Gamma$ being the electron Lorentz factor; see Sect.~4.2). Obviously, this effect cannot be accounted for by models neglecting the electron energy losses and assuming a homogeneous shell of emission (Chevalier \cite{che82b,che98}; P{\'e}rez-Torres et al. \cite{per01}; Weiler et al. \cite{wei02}; Soderberg et al. \cite{sod05}). 

However, significant deviations of the best-fit model from the data can be observed in Figure~\ref{fig10}. In particular, we see that the straight rising branches of the calculated light curves do not represent well the observations at 2 and 3.6~cm. This is most likely due to our treatment of external FFA (Sect.~2.2). It is possible that the structure of the CSM was more complicated at the time of explosion than that implicitly assumed by adopting an attenuation of the form $(1 - e^{-\tau_{\rm CSM}^{\rm clumps}}) / \tau_{\rm CSM}^{\rm clumps}$ with $\delta'=-3m$. This would be the case if, for example, the filling factor of clumpy material was not constant throughout the whole CSM. It is also possible that the CSM temperature $T_{\rm CSM}$ was not uniform but varied with radius. However, we note that the data at 6 and 20~cm are very well fitted. The best-fit normalization to the FFA optical depth is $K_3=3.0\times 10^4$, which gives from Eq.~(\ref{eq6}) $\dot{M}_{\rm RSG}=3.8 \times 10^{-5}~M_\odot$~yr$^{-1}$. 

Like for the brightness profile, the light curves provide clear evidence that the postshock magnetic field is essentially advected behind the shock. Indeed, we see in Figure~\ref{fig10} that the synchrotron flux declines much too rapidly after about day 100 in the model with damping of magnetic turbulence. 

The modeling of external FFA being uncertain, we now focus on the optically thin parts of the radio emission. In Figures~\ref{fig11} and \ref{fig12}, the flux density is multiplied by a time-dependent power law to set upright the decreasing parts of the light curves and the vertical scale is expanded. Figure~\ref{fig11} shows the effect of changing the magnetic field strength. For each value of $B_{u0}$, the electron injection rate was adjusted to provide a decent fit to the data at 3.6~cm. We see that only the model with $B_{u0}=50$~G represents the data at all the other wavelengths reasonably well. For example, for $B_{u0}=100$~G the calculated flux densities at 0.3~cm fall short of the data, whereas those at 20~cm are too high in the optically thin phase. This effect is due to the process of synchrotron cooling, whose rate increases with the magnetic field ($\tau_{\rm syn} \propto 1/B^2$) and which steepens the energy distribution of the accelerated electrons during their advection downstream. Thus, the mean propagated spectrum of nonthermal electrons is too steep (resp. too hard) for $B_{u0}=100$~G (resp. $B_{u0}=20$~G) to provide a good fit to the data at all frequencies. It is remarkable that when the synchrotron losses are taken into account, the degeneracy for the optically-thin emission between the magnetic field and the nonthermal electron density (e.g. Chevalier \cite{che98}) is lifted. 

Figure~\ref{fig12} shows the effect of changing $\eta_{\rm inj}^p$. The values of the electron injection parameter are somewhat arbitrary, except for the case $\eta_{\rm inj}^p=10^{-4}$. We see that in the test-particle case ($\eta_{\rm inj}^p=\eta_{\rm inj}^e=10^{-5}$), the decline of the optically-thin emission with time is too slow as compared to the data. This provides evidence that the pressure from accelerated ions is important to the structure of the blast wave. But we also see that for $\eta_{\rm inj}^p=3\times10^{-4}$ the light curves decline too rapidly, which shows that the shock modification is relatively weak. There is a partial correlation between $\eta_{\rm inj}^p$ and $B_{u0}$. By varying these two parameters and comparing the calculated curves to the data, I estimate the acceptable range for the proton injection rate to be $5 \times 10^{-5} < \eta_{\rm inj}^p < 2 \times 10^{-4}$. Higher values of $\eta_{\rm inj}^p$ are often used in studies of Galactic SNRs (e.g., Cassam-Chena\"i et al. (\cite{cas07}) adopted $\eta_{\rm inj}^p=10^{-3}$ for Tycho's remnant, based on the observed closeness of the blast wave and contact discontinuity). 

The time evolution of the synchrotron flux is related to the energy distribution of the emitting electrons (see, e.g., Chevalier~\cite{che82b}): the steeper the electron spectrum, the faster the decline of the optically-thin emission. As dicussed in Sect.~2.4, the main effect of the cosmic-ray pressure is to reduce the compression ratio of the subshock, $r_{\rm sub}$, whereas the overall compression ratio $r_{\rm tot}$ remains nearly constant. This shock modification affects essentially the particles of energies $<m_pc^2$ that remain in the vicinity of the subshock during the DSA process. As shown in Figure~\ref{fig3}, an increase of $\eta_{\rm inj}^p$ causes the energy distribution of the nonthermal electrons to steepens below $\sim$1 GeV, which is the relevant energy domain for the radio synchrotron emission. The light curves provide evidence that the spectral index of the shocked electrons in this energy range is slightly higher than the test-particle value $q_{\rm sub}=4$, which is effectively obtained with $\eta_{\rm inj}^p \approx 10^{-4}$. 

   \begin{figure*}[!]
   \centering
   \includegraphics[width=0.65\textwidth]{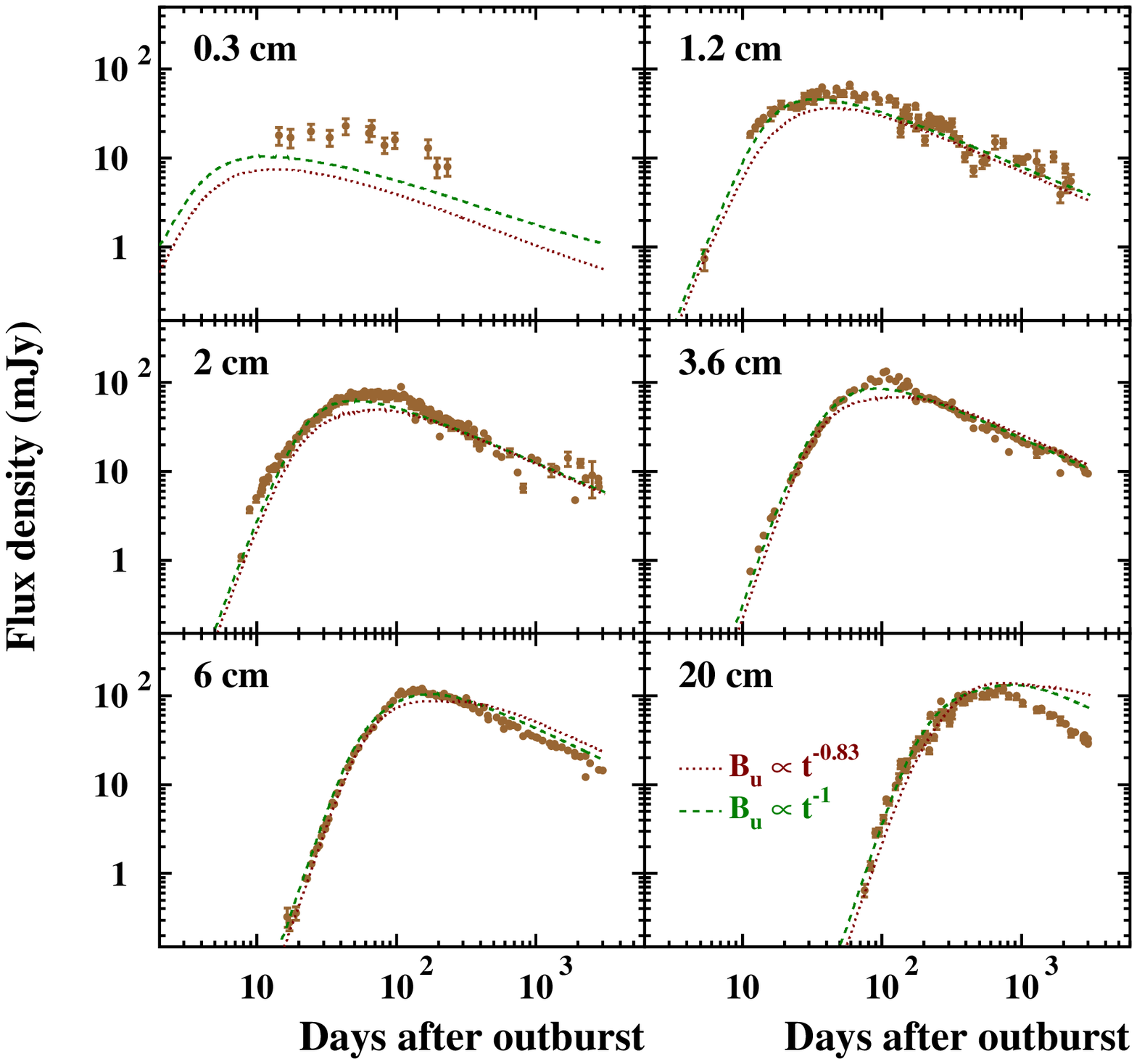}
      \caption{Same as Fig.~11 but for a CSM density profile of the form $\rho_{\rm CSM} \propto R^{-1.6}$. {\it Green dashed lines}: $B_u \propto t^{-1}$ and $\eta_{\rm inj}^p=2\times10^{-3}$; {\it red dotted lines}: $B_u \propto t^{-0.83}$ and $\eta_{\rm inj}^p=5\times10^{-3}$. For simplicity, FFA in the CSM is taken into account as before, with $\delta'=-3m$ and $K_3=3.0\times 10^4$. The other parameters are $B_{u0}=10$~G and $\eta_{\rm inj}^e=10^{-2}$.
           }
         \label{fig13}
   \end{figure*}

Although in most previous models of radio emission from SNe the shocked-electron spectral index is assumed to remain constant, a gradual steepening with time of the shocked-particle energy spectrum below $m_p c^2$ is to be expected. This is because of the decrease of the upstream magnetic field, which reduces the efficiency of gas heating via Alfv\'en wave damping in the precursor region and consequently reduces the subshock compression ratio. Thus, for $\eta_{\rm inj}^p=10^{-4}$, we find that $r_{\rm sub}$ decreases from 3.89 to 3.70 between day 10 and day 3100 after outburst (Fig.~\ref{fig4}), which causes $q_{\rm sub}$ to increase from 4.04 to 4.11 (Eq.~\ref{eq11}). 

For given values of $B_{u0}$ and $\eta_{\rm inj}^p$, the normalization of the optically-thin synchrotron flux is set by the product $\rho_u \times \eta_{\rm inj}^e \propto \dot{M}_{\rm RSG} \times \eta_{\rm inj}^e$. As the progenitor mass loss rate also fixes the level of FFA in the CSM, the electron injection rate can be uniquely determined. For $\eta_{\rm inj}^p=10^{-4}$, the best-fit parameter is $\eta_{\rm inj}^e=1.1\times10^{-5}$. To set the amplitude of the nonthermal electron spectrum in DSA models, several authors use the electron-to-proton density ratio at relativistic energies, $K_{\rm ep}$, rather than the electron injection rate (e.g., Ellison et al. \cite{ell00}). Here, we obtain at, e.g., 10~GeV $K_{\rm ep}=2.2\times10^{-3}$ at day 10 and $K_{\rm ep}=1.8\times10^{-3}$ at day 3100. The value of $K_{\rm ep}$ is higher at day 10 because $q_{\rm sub}$ is lower (see Eq.~\ref{eq10}). The  obtained values of $K_{\rm ep}$ are roughly consistent with those estimated for the blast wave of Tycho's SNR: $4\times10^{-3}$ (V\"olk et al.~\cite{vol02}), 
$\sim$1--$5 \times 10^{-3}$ (Cassam-Chena\"i et al. \cite{cas07}).  

\section{Discussion}

The model gives a consistent description of the large set of radio data available for SN~1993J. In particular, it is remarkable that the same magnetic field strength, $B_{u0}\sim 50$~G (in $B_u=B_{u0}(t/1~{\rm day})^{-1}$), is required to explain the high-resolution brightness profile measured by Bietenholz et al. (\cite{bie03}) at day 2787 after outburst and the set of radio light curves reported by Weiler et al. (\cite{wei07}). This further supports the assumption that the magnetic field at the shock varies approximately as $1/t$. Besides, both the brightness profile and the light curves show that the postshock magnetic field is not damped behind the blast wave but simply carried along by the plasma flow in the downstream region. We shall see in Sect.~4.3 that the obtained magnetic field evolution can be explained in terms of an amplification of the fluctuating component of the field by the Bell (\cite{bel04})'s nonresonant cosmic-ray streaming instability in the shock precursor region. But before, we are going to discuss in the light of the results two important assumptions of the model: the adopted CSM density profile (Sect.~4.1) and the relevant processes of cooling of the accelerated electrons (Sect.~4.2). 

\subsection{Circumstellar density profile and mass loss rate of the progenitor
star}

As already discussed in Sect.~2.2, there has been considerable debate about the structure of the CSM of SN~1993J. Several papers (Van Dyk et al. \cite{van94}; Fransson et al. \cite{fra96}; P{\'e}rez-Torres et al. \cite{per01}; Immler et al. \cite{imm01}; Weiler et al. \cite{wei07}) advocated for a flatter density profile, $\rho_{\rm CSM} \propto R^{-s}$ with $s=1.5$--1.7, than the $s=2$ power-law expected for a constant mass-loss rate and constant-velocity stellar wind prior to the SN explosion. On the other hand, Fransson \&  Bj\"ornsson (\cite{fra98}) obtained a good description of the radio data using the standard $s=2$ density profile. 

Figure~\ref{fig13} shows light curves calculated with $s=1.6$. The two sets of theoretical curves represent two different assumptions for the postshock magnetic field evolution: $B_d \propto t^{-1}$, consistent with the analysis of Sect.~2.3, and $B_d \propto t^{-0.83}$. The latter case, which is shown for completeness, can result from either a scaling of the magnetic energy density with the total postshock energy density ($B_d^2 \propto \rho_u V_s^2$; like in models 1 and 3 of Chevalier \cite{che96}) or a scaling of the postshock magnetic field with the pre-existing circumstellar field ($B_d \propto R^{-1}$; like in models 2 and 4 of Chevalier \cite{che96}). In order to reproduce the observed rate of decline of the radio emission in the optically thin phase, the spectrum of the emitting electrons has to be steeper for the flatter CSM density profile than for $s=2$. This is because $F_\nu \propto N_e \propto \rho_u$. It implies that $\eta_{\rm inj}^p$ should be higher for $s=1.6$ than for $s=2$. But we see that the resulting distributions of shocked electrons cannot provide simultaneously a good fit to the data at all wavelengths. The theoretical flux densities are clearly too low at 0.3 and 1.2~cm and too high at 6 and 20~cm as compared to the observations in the optically thin phase. This is somewhat similar to the case  of the too high magnetic field shown in Figure~\ref{fig11}. But with the low field strength $B_{u0}=10$~G adopted here, synchrotron losses are less important than Compton losses at early times (see Eq.~\ref{eq30}) and than adiabatic losses latter on. Therefore, reducing $B_{u0}$  below 10~G (and increasing $\eta_{\rm inj}^e$ at the same time ) does not help to improve the fit. Thus, we conclude that it is not possible to reproduce the observations  with $s=1.6$ in the framework of the present model. 
 
Weiler et al. (\cite{wei07}) recently adopted $s=1.61$ based on previous works and interpreted the flatter CSM density profile in terms of a steady decrease of the mass loss rate of the progenitor star, from $5.9 \times 10^{-6}~M_\odot$~yr$^{-1}$ $\sim$8000 years before explosion to 
$5.4 \times 10^{-7}~M_\odot$~yr$^{-1}$ at the time of the SN. These values are much lower than the mass loss rates estimated in the present as well as in previous works, which are in the range (2--6)$\times 10^{-5}~M_\odot$~yr$^{-1}$ just prior to  explosion (Van Dyk et al. \cite{van94}; Fransson et al. \cite{fra96}; Fransson \& Bj{\"o}rnsson \cite{fra98}; Immler et al. \cite{imm01}). 

As discussed in Appendix~A, the use of the fitting formalism developed by Weiler et al. (\cite{wei86,wei02} and references therein) may lead to uncertain physical parameters when both SSA and FFA in the CSM are important to explain the radio light curves. In this case, it could be justified to fix in the fitting procedure the time dependence of the mean magnetic field in the synchrotron-emitting region. As a check, I performed a simultaneous, least-squares fit to the light curve data for SN~1993J using the Weiler et al.'s formalism together with Eq.~(\ref{eqa13}), that is fixing $\delta''=\beta-2m-0.5$. I found the attenuation of the radio flux by the homogeneous component of the CSM to be not significant (i.e. $K_2$ compatible with zero) so I also fixed $A_{\rm CSM}^{\rm homog}=1$. I then obtained the best-fit parameters $K_1=5.4\times 10^3$, $\alpha=-0.79$, $\beta=-0.75$, $K_3=1.16\times 10^5$, $\delta'=-2.42$, and $K_5=6.9\times 10^4$, with $\chi^2=4250$ (for 552 degrees of freedom). The value of $K_3$ gives from Eq.~(\ref{eq6}) $\dot{M}_{\rm RSG}=7.6 \times 10^{-5}~M_\odot$~yr$^{-1}$, which is within a factor of two of the result obtained from the detailed model presented above. Noteworthy, the best-fit value of $\delta'$ is very close to the theoretical expectation: $\delta'=m(1-2s)=-2.49$ for $s=2$. This provides support to the assumption that the external free-free attenuation of the radio emission is mainly due to the inhomogeneous component of the CSM. 

\subsection{On the Coulomb energy losses of the nonthermal electrons}

The coulomb energy loss rate of relativistic electrons is given by (see Gould \cite{gou75})
\begin{equation}
{1 \over \tau_{\rm coul}} = {3 \sigma_{\rm T} c n_d^e \over 2\Gamma} F_{\rm coul} ~,
\label{eq32}
\end{equation}
where $F_{\rm coul}=\ln( \Gamma^{1/2} / \epsilon_p) +0.216$, $\sigma_{\rm T}$ is the Thompson cross section, $n_d^e$ is the number density of thermal electrons in the immediate postshock plasma, 
\begin{equation}
n_d^e = {\dot{M}_{\rm RSG} r_{\rm tot} \over 4 \pi R_s^2 u_w m_{\rm H}} \bigg({1+2X \over 1+4X}\bigg)~,
\label{eq33}
\end{equation}
and $\epsilon_p=7.27\times 10^{-17}n_d^e$ (in cgs units) is the normalized plasma energy. Coulomb cooling is more efficient than synchrotron cooling when $\tau_{\rm coul}<\tau_{\rm syn}$, the synchrotron energy loss rate being given by
\begin{equation}
{1 \over \tau_{\rm sync}} = {4 \sigma_{\rm T} \over 3 m_e c} {B_d^2 \over 8\pi} \Gamma~. 
\label{eq34}
\end{equation}

The typical Lorentz factor for the relativistic electrons radiating at frequency $\nu$ can be written as (see Eq.~\ref{eqb5}) 
\begin{equation}
\Gamma_c = 6.94 r_{\rm tot}^{-1/2} \bigg({B_{u0} \over 50~{\rm G}}\bigg)^{-1/2}  \bigg({\nu \over 8.4~{\rm GHz}}\bigg)^{1/2} \bigg({t \over 1{\rm~day}}\bigg)^{1/2}~, 
\label{eq35}
\end{equation}
where we have used for the magnetic field configuration across the shock $B_d \approx 0.83 r_{\rm tot} B_u$ (see Eqs.~16 and \ref{eq14}). Inserting $\Gamma_c$
into Eq.~(\ref{eq32}) and (\ref{eq34}), one obtains a limit on the time during which Coulomb cooling is more important than synchrotron cooling:
\begin{eqnarray}
t & < & \bigg[ 5.7 \bigg({R_0 \over 3.49 \times 10^{14}{\rm~cm}}\bigg)^{-2}  \bigg({\dot{M}_{\rm RSG} \over 3.8 \times 10^{-5}~M_\odot{\rm~yr}^{-1}}\bigg) \nonumber \\ 
& \times & \bigg({u_w \over 10{\rm~km~s}^{-1}}\bigg)^{-1}  \bigg({B_{u0} \over 50~{\rm G}}\bigg)^{-1} \bigg({\nu \over 8.4~{\rm GHz}}\bigg)^{-1} \bigg]^{1/(2m-1)}~{\rm days}.
\label{eq36}
\end{eqnarray}
Here, we have used $F_{\rm coul} \approx 30$ and $X=0.3$. This result shows that Coulomb energy losses can be safely neglected for frequencies $\nu \geq 8.4$~GHz. They could play a role, however, during the rising phase of the light curve at 1.4~GHz (20~cm), for $\sim 210$~days after outburst. But at that time, the 20~cm emission was attenuated by SSA and the emitted flux density depended only weakly on the propagated spectrum of nonthermal electrons. Thus, the neglect of Coulomb cooling in the model is justified. 

\subsection{Magnetic field amplification}

Fransson \&  Bj\"ornsson (\cite{fra98}) modeled the radio emission of SN~1993J assuming that the radiation is emitted from a shell of uniform magnetic field. They found $\langle B \rangle \approx 340 (t/1~{\rm day})^{-1}$~G (or alternatively $\langle B \rangle \approx 64 (R_s/10^{15}{\rm~cm})^{-1}$~G), which is in reasonable agreement with both our first estimate (Eq.~\ref{eq7p}) and the mean postshock magnetic field resulting from our detailed analysis. Our best-fit value for the immediate postshock magnetic field is $B_d \approx 0.83 r_{\rm tot} B_u \approx  180 (t/1~{\rm day})^{-1}$~G, given $r_{\rm tot} \approx 4.3$ (Fig.~\ref{fig4}). The mean magnetic field strength in the synchrotron-emitting region is expected to be higher than $B_d$ by a factor of less than two, given that the field is mainly advected in the downstream plasma (see Fig.~\ref{fig5}), i.e. $180 < \langle B \rangle (t/1~{\rm day}) < 360$~G. We note, however, that in our detailed model all the emitting electrons do not see the same mean magnetic field, because due to the energy losses the downstream transport of the nonthermal particles depends on the energy they acquired at the shock front. 

P{\'e}rez-Torres et al. (\cite{per01}) were able to obtain a good fit to the radio light curves with $\langle B \rangle \approx 56 (t/1~{\rm day})^{-0.86}$~G. But they did not take into account the electron energy losses and assumed the existence of a low-energy cutoff in the distribution of the relativistic particles, whose main effect is to flatten the radio light curves at the shortest wavelengths. But as discussed in Sect.~3.2, the observed flattening of the light curves at early epochs is naturally explained by the strong synchrotron losses suffered by the radiating electrons. Moreover, a low-energy cutoff in the particle distribution function is not supported by the DSA theory. 

As already pointed out by Fransson \&  Bj\"ornsson (\cite{fra98}) and P{\'e}rez-Torres et al. (\cite{per01}) the  magnetic field inferred from the radio observations is at least two orders of magnitude higher than the magnetic field of stellar origin that is expected to preexist in the wind of the red supergiant progenitor. The high magnetic field in the blast wave region is likely a result of an amplification associated with the production of nonthermal particles by nonlinear DSA. The diffusive streaming of cosmic-rays in the upstream plasma is expected to produce strong turbulence, which could amplify the chaotic component of the magnetic field (e.g., Bell \& Lucek \cite{bel01}; Amato \& Blasi \cite{ama06}; Vladimorov et al. \cite{vla06}). Two different mechanisms of turbulence generation are discussed for SNR shocks. First, due to the cosmic-ray anisotropy in the shock precursor, a resonant instability is predicted to exist in this region and to produce a rapid and intense excitation of Alfv\'en waves (e.g., McKenzie \& V\"olk \cite{mck82}; Bell \& Lucek \cite{bel01}). Bell (\cite{bel04}) also found a nonresonant streaming instability caused by the Lorentz force associated with the cosmic-ray electric current, which could strongly amplified MHD perturbations of short wavelengths. The overall magnetic field amplification could be the result of both instabilities operating in the shock precursor. 

   \begin{figure}
   \centering
   \includegraphics[width=0.45\textwidth]{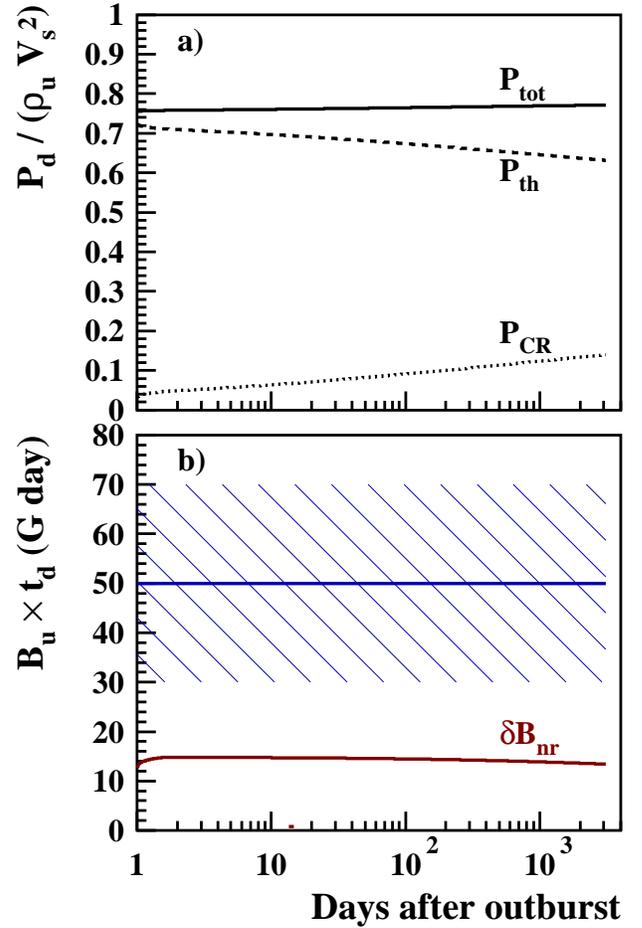}
      \caption{{\bf (a)} Normalized immediate postshock pressures in thermal ($P_{\rm th}$) and nonthermal ($P_{\rm CR}$) particles ($P_{\rm tot}=P_{\rm th}+P_{\rm CR}$) as a function of time after outburst, for the best obtained solution ($\dot{M}_{\rm RSG}=3.8 \times 10^{-5}~M_\odot$~yr$^{-1}$, $B_{u0}=50$~G, $\eta_{\rm inj}^p=10^{-4}$, and $\eta_{\rm inj}^e=1.1\times10^{-5}$). {\bf (b)} Magnetic field immediately upstream from the subshock $B_u$ times $(t/1~{\rm day})$. The hatched area shows the best-fit value obtained from the radio data, $B_u=B_{u0}(t/1~{\rm day})^{-1}$ with $B_{u0}=50\pm20$~G, and the red solid curve shows the fluctuating magnetic field expected from the nonresonant instability (eq.~\ref{eq37}).}
         \label{fig14}
   \end{figure}

The saturation value of the random magnetic field amplified by the nonresonant instability, $\delta B_{\rm nr}$, can be estimated from equation~(28) of  Pelletier et al. (\cite{pel06}; see also Bell \cite{bel04}):
\begin{equation}
{\delta B_{\rm nr}^2 \over 8 \pi} \cong {3 V_s P_{\rm CR} \over 2 \Phi c}~, 
\label{eq37}
\end{equation}
where $\Phi=\ln(p_{\rm max}^p/m_pc) \approx 15$ (see Sect.~4.4 below) and $P_{\rm CR}$ is the cosmic-ray pressure at the shock front. The normalized pressure $\xi_{\rm CR}=P_{\rm CR}/\rho_u V_s^2$ calculated with the nonlinear DSA model from the best-fit parameters is shown in Figure~\ref{fig14}a. We see that $\xi_{\rm CR}$ increases with time, but is always much lower than the normalized postshock thermal pressure. For such a weakly cosmic-ray-modified shock, we have in good approximation $\xi_{\rm CR} \propto p_{\rm inj}^p / V_s^2$ (see Berezhko \& Ellison \cite{ber99}, eq.~34), which leads to $\xi_{\rm CR} \propto V_s^{-1}$ given that $p_{\rm inj}^p \propto p_{\rm th} \propto V_s$ (Blasi et al. \cite{bla05}). Here, $p_{\rm th}$ is the most probable momentum of the thermal protons in the postshock plasma. We then obtain from Eq.(\ref{eq37}) 
\begin{equation}
\delta B_{\rm nr} \propto \big(\xi_{\rm CR} \rho_u V_s^3\big)^{1/2}  \propto t^{-1}~. 
\label{eq38}
\end{equation}
As previously stated (Sect.~2.3) the magnetic field evolution expected from the nonresonant instability mechanism uncovered by Bell (\cite{bel04}) is consistent with that inferred from the data. As shown in Figure~\ref{fig14}b,  the strength of the field amplification calculated from Eq.~(\ref{eq37}) is also quite close to that required to explain the radio flux.  

Following Pelletier et al. (\cite{pel06}) the relative importance of the resonant and nonresonant regimes of the streaming instability can be estimated from 
\begin{equation}
{\delta B_{\rm res}^2 \over \delta B_{\rm nr}^2} \sim \bigg({\xi_{\rm CR} c \over V_s}\bigg)^{1/2}~. 
\label{eq39}
\end{equation}
This expression shows that in SN~1993J the nonresonant mechanism is expected to dominate at early times, but the resonant instability becomes more and more important as the shock is slowing down. 

According to Bell (\cite{bel04}) the minimum timescale for growth of MHD waves driven by the nonresonant instability at a distance $z$ upstream from the subshock is given by
\begin{equation}
\tau_{\rm nr}(z) \approx {\Phi c E_p(z) \over e \epsilon_{\rm nt} V_s^3 (4 \pi \rho_u)^{0.5}}~, 
\label{eq39b}
\end{equation}
where $-e$ is the electronic charge, $\epsilon_{\rm nt}$ is the cosmic-ray acceleration efficiency (see Figure~\ref{fig4}) and $E_p(z) \approx 3 z e B_u V_s c^{-1} \eta_{\rm mfp}^{-1}$ is the energy of the accelerated protons whose upstream diffusion length is equal to $z$. Numerically, the MHD growth timescale for SN~1993J is
\begin{eqnarray}
\tau_{\rm nr}(z) \approx 3.3 \times 10^{-2} \bigg({\Phi \over 15}\bigg) \bigg({\epsilon_{\rm nt} \over 0.1}\bigg)^{-1} \bigg({E_p(z) \over 10^{15}{\rm~eV}}\bigg) \bigg({t \over 1{\rm~day}}\bigg)^{3-2m}{\rm~day}. 
\nonumber \\ 
\label{eq39t}
\end{eqnarray}
Thus, we have $\tau_{\rm nr} \ll t$  for $E_p(z)\leq 10^{15}$~eV. This shows that the turbulent magnetic field amplified by the nonresonant instability should reach its saturation value over most of the shock precursor region soon after the onset of particle acceleration.

Because the damping of MHD turbulence can efficiently heat the thermal gas in the shock precursor region, a high amplification of the fluctuating magnetic field could at the same time provide a strong limitation for the modification of the shock structure induced by the cosmic-ray pressure. According to, e.g., Berezhko \& Ellison (\cite{ber99}) the degree of heating of the shock precursor via Alfv\'en wave damping is set by the ratio $M_{S,u}^2 / M_{A,u}$ ($M_{S,u}$ and $M_{A,u}$ are the upstream sonic and Alfv\'en Mach numbers, respectively; see Sect.~2.4). Identifying $B_u$ and $\delta B_{\rm nr}$ one obtains from Eq.~(\ref{eq38}) the scaling relation  
\begin{equation}
{M_{S,u}^2  \over M_{A,u}} \propto  \xi_{\rm CR}^{1/2} V_s^{5/2}~, 
\label{eq40}
\end{equation}
which shows that for the same cosmic-ray pressure the shock modification should be less pronounced in a high-speed shock than in a slower one. 

Bj\"ornsson \& Fransson (\cite{bjo04}) modeled the radio and X-ray emission from SN~2002ap with an accelerated electron energy index $\gamma \approx 2$, as expected for a test-particle shock. The above scaling relation suggests that the absence of clear shock modification in this object is a consequence of the very high velocity of the blast wave ($V_s \sim 7 \times 10^4$~km~s$^{-1}$; Bj\"ornsson \& Fransson \cite{bjo04}). 

Various studies of the synchrotron emission produced in young SNRs led Berezhko \& V\"olk (\cite{ber06}; see also Berezhko~\cite{ber08}) to propose the following empirical relation between the amplified magnetic field energy and the cosmic-ray pressure:
\begin{equation}
{B_u^2 \over 8 \pi} \approx 5 \times 10^{-3} P_{\rm CR}~. 
\label{eq41}
\end{equation}
For SN~1993J, a better relation would be (see Eq.~\ref{eq37}) 
\begin{equation}
{B_u^2 \over 8 \pi} \approx 10^{-1} P_{\rm CR} \bigg({V_s \over3 \times 10^4{\rm~km~s}^{-1}}\bigg)~. 
\label{eq42}
\end{equation}
It is remarkable that this expression could provide a fair estimate of the magnetic field amplification in the high-speed blast wave of SN~1993J ($V_s/c \sim 0.1$), as well as in slower shocks of Galactic SNRs, which are at the end of the free expansion phase of the post-SN evolution or in the Sedov phase ($V_s/c \sim 0.01$). In these objects, however, the resonant regime of the cosmic-ray streaming instability is expected to dominate the field amplification. 

\subsection{SN~1993J and the origin of cosmic rays}

   \begin{figure}
   \centering
   \includegraphics[width=0.45\textwidth]{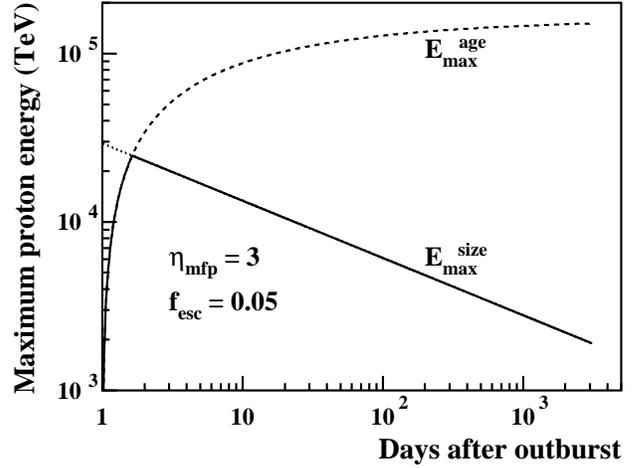}
      \caption{Maximum energy of shock-accelerated protons as a function of time after outburst. $E_{\rm max}^{\rm age}$ and $E_{\rm max}^{\rm size}$ are the maximum energies caused by the finite shock age and size, respectively. The true maximum proton energy ({\it solid line}) is the minimum of these two quantities.}
         \label{fig15}
   \end{figure}

As already pointed out by Bell \& Lucek (\cite{bel01}) cosmic-rays were probably accelerated to very high energies in the blast wave of SN~1993J short after shock breakout. Calculated maximum proton energies are shown in Figure~\ref{fig15}. The quantity $E_{\rm max}^{\rm time}$ is the maximum energy caused by the finite age of the shock (Baring et al. \cite{bar99}; Tatischeff \& Hernanz \cite{tat07}) and is obtained by time integration of the DSA rate from an initial acceleration time assumed to be $t_0=1$~day after outburst. The maximum energy $E_{\rm max}^{\rm size}$ is caused by the finite spatial extend of the shock and is calculated by equalling the upstream diffusion length of the accelerated protons to some fraction $f_{\rm esc}$ of the shock radius. Numerically, we find
(see, e.g., Tatischeff \& Hernanz \cite{tat07})
\begin{eqnarray}
E_{\rm max}^{\rm size} & = & 1.75 \times 10^{18} \bigg({f_{\rm esc} \over \eta_{\rm mfp}}\bigg) \bigg({m \over 0.83}\bigg)^{-1} \bigg({B_{u0} \over 50~{\rm G}}\bigg) \nonumber \\ 
& \times & \bigg({V_0 \over 3.35 \times 10^4{\rm~km~s}^{-1}}\bigg)^2 \bigg({t \over 1{\rm~day}}\bigg)^{2(m-1)}~~{\rm eV}. 
\label{eq43}
\end{eqnarray}
We see in Figure~\ref{fig15} that for $f_{\rm esc}=0.05$ the size limitation of the shock becomes rapidly more restrictive than the age limitation. This conclusion is independent of $t_0$. However, the value of $f_{\rm esc}$, which depends on the upstream turbulence generated by the accelerated particle streaming (Marcowith et al. \cite{mar06}), is uncertain\footnote{A simple estimate of $f_{\rm esc}$ may be derived by assuming that the upstream escape boundary at $z=f_{\rm esc}R_s$ corresponds to the distance for which the MHD growth timescale $\tau_{\rm nr}(z)=t$. We then obtain from Eq.~(\ref{eq39b})
\begin{equation}
f_{\rm esc} \approx {\epsilon_{\rm nt} \eta_{\rm mfp} m M_{A,u} \over 3 \Phi}~. 
\label{eqfootnote}
\end{equation}
For $\epsilon_{\rm nt}=0.1$, $\Phi=15$, $\eta_{\rm mfp}=3$, $m=0.83$, and $M_{A,u}=9.5$ (see above), we get $f_{\rm esc}$$\approx$ 0.05. However, the full interpretation of this result is beyond the scope of this paper.}. But in any case, the resulting maximum proton energy (i.e. the minimum of $E_{\rm max}^{\rm size}$ and $E_{\rm max}^{\rm age}$) is lower than the previous estimate of Bell \& Lucek (\cite{bel01}), $E_{\rm max}^p \sim 3\times10^{17}$~eV. Moreover, we did not take into account the nonlinear effects recently pointed out by Ellison \& Vladimirov (\cite{ell08}), which could further reduce $E_{\rm max}^p$. However, it is likely that SN~1993J has accelerated protons above $3\times10^{15}$~eV, i.e. the energy of the spectral "knee" above which the all-particle cosmic-ray spectrum measured near Earth shows a significant steepening. This could provide support to the scenario first proposed by V\"olk \& Biermann (\cite{vol88}) that the very-high energy Galactic cosmic-rays are produced by massive stars exploding into their former stellar wind. 

In this context, it is instructive to estimate the total energy acquired by the cosmic-ray particles in SN~1993J during the early stage of interaction between the SN ejecta and the red supergiant wind:
\begin{equation}
E_{\rm CR} \cong \int_{t_0}^{t_f} \epsilon_{\rm nt} \times {1 \over 2} 
\rho_u V_s^3 \times 4\pi R_s^2 dt = 7.4 \times 10^{49}~\rm erg,
\label{eq44}
\end{equation}
where $t_f=3100$~days and $\epsilon_{\rm nt}$ is the acceleration efficiency shown in Figure~\ref{fig4}. It is remarkable that the value obtained for $E_{\rm CR}$ is very close to the mean energy per SN required to account for the Galactic cosmic-ray luminosity, $\approx 7.5 \times 10^{49}$~erg (Tatischeff \cite{tat08}). Before day $\sim$3100, i.e. during the first $\sim$8.5~years after explosion, the forward shock processed in the expansion a total energy of $4.0 \times 10^{50}$~erg and the mean acceleration efficiency was $\langle \epsilon_{\rm nt} \rangle=19\%$. 

Most of the cosmic rays accelerated during this early stage were advected downstream from the shock and will remain inside the SNR until it merges with the interstellar medium. These particles suffer adiabatic losses and their energy decreases like $R_s^{-1}$ (e.g. Drury \cite{dru96}). Thus, they will make a negligible contribution to the interstellar cosmic-ray population at the end of the SNR. But as shown in Figure~\ref{fig15}, the highest-energy cosmic-rays are expected to escape continuously the shock system from upstream, due to the finite size of the acceleration region. The value obtained for $E_{\rm CR}$ (Eq.~\ref{eq44}) suggests that these very-high energy particles could be important to the origin of interstellar cosmic rays above $\sim 10^{15}$~eV.  

At day $\sim$3100 after outburst, the blast wave started to expand into a more diluted CSM resulting from a lower mass loss rate of the SN progenitor more than $\sim$9000 years before explosion (see Weiler et al. \cite{wei07}). The CSM density structure of SN~1993J can be explained by stellar evolution calculations that show that red supergiant stars can undergo a stage of intensified mass loss for few 10$^4$~years before SN explosion (Heger et al. \cite{heg97}). After day $\sim$3100 post-outburst, an additional number of very-high energy cosmic-rays could have escaped the shock region into the interstellar medium, because their diffusion length $\ell_u$ has increased ($\ell_u \propto B_u^{-1} \propto \rho_u^{-1/2}$; see eq.~\ref{eq38}). Significant variations in the mass loss activity of red supergiant stars prior to explosion are commonly inferred from radio observations of type II SNe (Weiler et al. \cite{wei02}).

These results suggest that Galactic cosmic rays of energies above $\sim 10^{15}$~eV could be produced immediately after the explosion of massive stars (V\"olk \& Biermann \cite{vol88}; Biermann \cite{bie93}), provided that the acceleration efficiency found for SN~1993J is typical of most SNe and that the mass-loss rate obtained for the progenitor of this type IIb SN, $\dot{M}_{\rm RSG}=3.8 \times 10^{-5}~M_\odot$~yr$^{-1}$, is close to the mean mass-loss rate of red supergiant stars prior to explosion. 

\subsection{SN~1993J as a gamma-ray source}

Kirk et al. (\cite{kir95}) pointed out that radio SNe could emit a large flux of very-high energy gamma rays arising from the decay of neutral pions created in hadronic collisions of cosmic rays with target nuclei. A detection of the predicted TeV emission with ground-based atmospheric Cherenkov telescopes would provide a direct confirmation of efficient acceleration of high-energy protons in SN explosions. 

According to Kirk et al. (\cite{kir95}), SN~1993J was a promising target for detection, with a predicted flux of gamma-rays above 1 TeV of $F_\gamma(>1~{\rm TeV})=2 \times 10^{-12}$~cm$^{-2}$~s$^{-1}$ for years after outburst. But their calculations assumed a much flatter CSM density profile, $\rho_{\rm CSM} \propto R^{-1.5}$, than the standard radial dependence $\rho_{\rm CSM} \propto R^{-2}$ found here. 

The gamma-ray flux can be estimated from 
\begin{equation}
F_\gamma(>1~{\rm TeV}) \approx {q_\gamma(>1~{\rm TeV}) \langle n_p \rangle
\langle \epsilon_{\rm CR} \rangle V \over 4 \pi D^2},
\label{eq45}
\end{equation}
where $q_\gamma(>1~{\rm TeV}) \approx 10^{-17}$~photons~s$^{-1}$~erg$^{-1}$~cm$^3$~(H-atom)$^{-1}$ is the gamma-ray emissivity, which depends on the cosmic-ray spectrum produced at the shock (Drury et al. \cite{dru94}) and $\langle n_p \rangle \approx n_d^e/(1+2X)$ (see eq.~\ref{eq33}) is the mean density of hydrogen nuclei in the downstream volume $V$ occupied by the cosmic rays, whose mean energy density is $\langle \epsilon_{\rm CR} \rangle \approx 3 P_{\rm CR}$. In first approximation we can identify the volume $V$ with that of the shocked CSM:
\begin{equation}
V \approx {4 \pi \over 3} \big( R_s^3 -R_{\rm CD}^3 \big) \approx {2 \pi \over 3} R_s^3~,
\label{eq46}
\end{equation}
where $R_{\rm CD}=R_s/1.27$ is the radius of the contact discontinuity (Fig.~\ref{fig5}). We then obtain
\begin{eqnarray}
F_\gamma(>1~{\rm TeV}) & \approx & 2 \times 10^{-12} \bigg({D \over 3.63{\rm~~Mpc}}\bigg)^{-2} \bigg({\dot{M}_{\rm RSG} \over 3.8 \times 10^{-5}~M_\odot{\rm~yr}^{-1}}\bigg)^2 \nonumber \\ 
& \times & \bigg({u_w \over 10{\rm~km~s}^{-1}}\bigg)^{-2} \bigg({t \over 1{\rm~day}}\bigg)^{-1}
~~{\rm cm}^{-2}~{\rm s}^{-1},
\label{eq47}
\end{eqnarray}
where we have used $\xi_{\rm CR} \approx 0.04 (t/1~{\rm day})^{1-m}$ (see Fig.~\ref{fig14}).

But the TeV gamma-ray emission can be attenuated by the pair production process  $\gamma + \gamma \rightarrow e^+ + e^-$ in the strong radiation field from the SN ejecta. The corresponding optical depth at the gamma-ray energy $E_\gamma$ can be estimated from
\begin{equation}
\tau_{\gamma\gamma}(E_\gamma) \approx R_s \kappa_{\gamma\gamma}(E_\gamma)~,
\label{eq48}
\end{equation}
where 
\begin{equation}
\kappa_{\gamma\gamma}(E_\gamma) = {45 \sigma_{\rm T} U_{\rm rad} \over 8 \pi^4 k T_{\rm bb}}f_{\gamma\gamma}(E_\gamma,T_{\rm bb})
\label{eq49}
\end{equation}
is the absorption coefficient in the diluted blackbody radiation field of temperature $T_{\rm bb}$ and energy density at the shock $U_{\rm rad}$  (Gould \& Schr\'eder \cite{gou67}). The SN ejecta emitted mainly in the UV band during the first week after explosion and then in the optical for more than 120 days with a mean blackbody temperature $T_{\rm bb} \approx 7000$~K (Richmond et al. \cite{ric94}; Lewis et al. \cite{lew94}). For this temperature, the function $f_{\gamma\gamma}(E_\gamma,T_{\rm bb})$ reaches its maximum value $\approx 1$ at $E_\gamma \approx 1$~TeV (see Gould \& Schr\'eder \cite{gou67}). Using $U_{\rm rad} \approx L_{\rm bol} /(4 \pi c R_s^2)$ with the power-law approximation of Fransson \& Bj\"ornsson (\cite{fra98}) for the evolution of the SN bolometric luminosity (Eq.~\ref{eq29}), we obtain
\begin{equation}
\tau_{\gamma\gamma}(E_\gamma) \approx 179 \bigg({t \over 10~{\rm days}}\bigg)^{-1.73}~.
\label{eq50}
\end{equation}
Thus, the early TeV emission of SN~1993J was strongly attenuated by photon-photon absorption. Multiplying the unabsorbed flux $F_\gamma(>1~{\rm TeV})$ given by Eq.~(\ref{eq47}) by the attenuation factor $A_{\gamma\gamma}=\exp(-\tau_{\gamma\gamma})$, we find that the flux reached a maximum of $\approx$$4 \times 10^{-15}$~cm$^{-2}$~s$^{-1}$ about 270 days after explosion, which is a factor of $\approx$40 lower than the 5$\sigma$ sensitivity of the High Energy Stereoscopic System (H.E.S.S.) of atmospheric Cherenkov telescopes for 25~hours of observation near zenith (Aharonian et al. \cite{aha06}). 

Below $\sim$50~GeV, gamma rays are not attenuated by pair production in collisions with optical photons. The total gamma-ray flux in the energy range 1--50~GeV can be readily estimated from Eq.~(\ref{eq45}) using the emissivity $q_\gamma(>1~{\rm GeV}) \approx 10^{-14}$~photons~s$^{-1}$~erg$^{-1}$~cm$^3$~(H-atom)$^{-1}$ (Drury et al. \cite{dru94}). It gives for SN~1993J 
\begin{equation}
F_\gamma(>1~{\rm GeV}) \approx 2 \times 10^{-9} \bigg({t \over 1{\rm~day}}\bigg)^{-1}~~{\rm cm}^{-2}~{\rm s}^{-1}. 
\label{eq51}
\end{equation}
The mean flux of gamma rays above 1~GeV from day 1 to day 8 after outburst is then $\approx$6$\times 10^{-10}$~cm$^{-2}$~s$^{-1}$, which is $\approx$14 times lower than the 5$\sigma$ sensitivity of the LAT instrument on the {\it Fermi} Gamma-ray Space Telescope for one week of observation in the survey mode\footnote{As calculated with the LAT source detectability tool at http://fermi.gsfc.nasa.gov/ssc/proposals/detectability.html.}. This suggests that a SN similar to SN~1993J in terms of progenitor wind properties and cosmic-ray acceleration efficiency could be detected in $\pi^0$-decay gamma-rays with {\it Fermi} out to a maximum distance of $\sim$1~Mpc. 

\section{Conclusions}

Inspired by previous works on the morphology of synchrotron emission in Galactic SNRs (Cassam-Chena\"i et al. \cite{cas05}), I have employed a model to explain the radio emission of SN~1993J, which includes nonlinear effects of the diffusive shock acceleration mechanism. The model states that cosmic-ray ions are accelerated at the SN blast wave and their pressure modify the shock structure. Therefore, the energy spectrum of the accelerated electrons depends on the rate of injection of shocked protons into the acceleration process. The nonthermal electrons suffer both adiabatic and radiative losses during their advection downstream the shock, which modify their energy distribution. The hydrodynamic evolution of the postshock plasma is calculated by assuming that the SN expansion is self-similar. The relativistic electrons emit synchrotron radiation in the ambient magnetic field and the corresponding radio emission is worked out from radiative transfer calculations that include the process of synchrotron self-absorption. The magnetic field amplification associated with the efficient production of cosmic-rays is not calculated in the framework of the model, but instead the strength of the amplified field is determined from a fit to the radio data. The model contains three other parameters to be fitted to the data: the shocked proton and electron injection rates into the DSA process and the mass loss rate of the SN progenitor.  

The extensive radio observations of SN~1993J make this object a unique "laboratory" to study cosmic-ray acceleration in a SN shock. In particular, the VLBI imaging observations that resulted in detailed measurements of the SN expansion are of prime interest for the DSA theory. By applying the model to both the high-resolution brigthness profile at 3.6~cm measured by Biethenholz et al. (\cite{bie03}) and radio light curves at six frequencies reported by Weiler et al. (\cite{wei07}), I have obtained the following main conclusions:
\begin{enumerate}
\item
The CSM density profile is consistent with a constant, steady wind of the red supergiant progenitor of the SN for $\sim$9000~years before explosion. Contrary to previous claims, SN~1993J is not markedly different from the other studied radio SNe with regards to the mass loss evolution of the progenitor star. The best-fit mass-loss rate is $\dot{M}_{\rm RSG}=3.8 \times 10^{-5}~M_\odot$~yr$^{-1}$ for the assumed stellar wind terminal velocity $u_w=10$~km~s$^{-1}$ and CSM temperature $T_{\rm CSM}=2 \times 10^5$~K.
\item
The observed morphology of the radio emission provides evidence that part of the synchrotron radiation ($\approx$17\% of the total flux density at day 2787 post-outburst) is produced by electrons accelerated at the reverse shock.  
\item
The best-fit rate of injection of cosmic-ray protons into the DSA process at the forward shock is $\eta_{\rm inj}^p=10^{-4}$, with an estimated error of a factor of two. The calculated fraction of the total incoming energy flux converted by the blast wave to cosmic-ray energy during the first $\sim$8.5~years after explosion is $\langle \epsilon_{\rm nt} \rangle=19\%$. The shock modification induced by the backpressure from the accelerated ions is increasing with time after outburst. However, the cosmic-ray pressure accounts for less than 20\% of the total postshock pressure at all times, such that the shock modification remains relatively weak. 
\item 
The best-fit cosmic-ray electron injection rate is $\eta_{\rm inj}^e=1.1 \times 10^{-5}$ and the calculated electron-to-proton density ratio at 10~GeV is $K_{\rm ep} \approx 2 \times 10^{-3}$. The obtained value of $K_{\rm ep}$ is lower than that measured in the Galactic cosmic-rays near Earth, $\approx$10$^{-2}$. The latter number, however, results from cosmic-ray transport in the Galaxy and therefore is probably not indicative of the relativistic electron-to-proton ratio at the source of cosmic-ray acceleration. 
\item
The synchrotron energy losses suffered by the radiating electrons in the postshock magnetic field are important for the modeling of both the radio light curves and the morphology of the radio emission. The inferred magnetic field is broadly consistent with that expected from an amplification in the shock precursor region by the nonresonant regime of the cosmic-ray streaming instability. The magnetic field immediately upstream from the subshock is found to be $B_u=B_{u0}(t/1~{\rm day})^{b}$ with $b \approx -1$ and $B_{u0}=50 \pm 20$~G. The measured field strength is higher by a factor of 2 to 5 than the saturated value of the turbulent magnetic field predicted from the model of Bell (\cite{bel04}) for the turbulence generation. The energy density in the amplified magnetic field is found to obey the relation
\begin{equation}
{B_u^2 \over 8 \pi} \approx 10^{-1} P_{\rm CR} \bigg({V_s \over3 \times 10^4{\rm~km~s}^{-1}}\bigg)~. 
\label{eq52}
\end{equation}
\item
The turbulent magnetic field amplified in the precursor region is not damped behind the shock (as proposed by Pohl et al. \cite{poh05} for Galactic SNRs) but essentially carried along by the plasma flow in the downstream region. 
\item
The magnetic field amplification increases the DSA rate, thus allowing the rapid acceleration of cosmic-ray protons to energies well above 10$^{15}$~eV. The proton maximum energy is find to be limited by the finite confinement size of the shock, which implies that the highest-energy cosmic rays continuously escaped the acceleration region from upstream during the first $\sim$8.5~years after explosion. After that time, when the blast wave passed the limit of the dense CSM established by a stage of high mass loss from the red supergiant progenitor prior to explosion, an additional number of very-high energy protons probably escaped into the interstellar medium. The results obtained for this SN provide support to the model of V\"olk \& Biermann (\cite{vol88}) and Biermann (\cite{bie93}) that the Galactic cosmic-rays above $\sim$10$^{15}$~eV are accelerated in the explosion of massive stars, during the early stage of interaction of the SN ejecta with the progenitor wind. 
\item
The early emission from SN~1993J of TeV gamma-rays produced in hadronic collisions of accelerated cosmic-rays with ambient material was strongly attenuated by pair production in the dense radiation field from the SN ejecta. The flux at Earth of photons of energy above 1~TeV reached a maximum of only $\approx$$4 \times 10^{-15}$~cm$^{-2}$~s$^{-1}$ at day $\sim$270 after explosion. Above 1~GeV, the flux is found to be $F_\gamma(>1~{\rm GeV}) \approx 2 \times 10^{-9} (t / 1{\rm~day})^{-1}$~cm$^{-2}$~s$^{-1}$. This result suggests that type II SNe could be detected in $\pi^0$-decay gamma-rays with the {\it Fermi} Gamma-ray Space Telescope out to a maximum distance of only $\sim$1~Mpc. 
\end{enumerate}

\begin{acknowledgements}

It is a pleasure to thank Margarita Hernanz for numerous discussions and her hospitality at IEEC-CSIC, where most of this work has been done. I am greatly indebted to Jean-Pierre Thibaud for his generous and stimulating inputs throughout the writing of the paper. I am also greatful to Gamil Cassam-Chena\"i, Anne Decourchelle, J\"urgen Kiener, Alexandre Marcowith, and R\'egis Terrier for numerous illuminating discussions and to Roger Chevalier for his critical reading of the manuscript. Financial support from the Generalitat de Catalunya through the AGAUR grant 2006-PIV-10044 and the project SGR00378 is acknowledged.

\end{acknowledgements}

\appendix

\section{Magnetic field evolution from the parametric model of Weiler et al.}

Weiler et al. (\cite{wei86,wei02} and references therein) have developed a semi-phenomenological model to described the light curves of radio SNe. In this model, the flux density at a given frequency $\nu$ and time $t$ after outburst can be expressed as
\begin{equation}
F({\rm mJy})= K_1 \bigg({\nu \over 5{\rm~GHz}}\bigg)^\alpha \bigg({t \over 1{\rm~day}}\bigg)^\beta A_{\rm CSM}^{\rm homog} A_{\rm CSM}^{\rm clumps} A_{\rm SSA}~,
\label{eqa1}
\end{equation}
where
\begin{equation}
A_{\rm CSM}^{\rm homog}=\exp(-\tau_{\rm CSM}^{\rm homog})~,
\label{eqa2}
\end{equation}
\begin{equation}
A_{\rm CSM}^{\rm clumps} = {1 - \exp(-\tau_{\rm CSM}^{\rm clumps}) \over \tau_{\rm CSM}^{\rm clumps}}~,
\label{eqa3}
\end{equation}
and
\begin{equation}
A_{\rm SSA} = {1 - \exp(-\tau_{\rm SSA}) \over \tau_{\rm SSA}}~,
\label{eqa4}
\end{equation}
with
\begin{equation}
\tau_{\rm CSM}^{\rm homog} = K_2 \bigg({\nu \over 5{\rm~GHz}}\bigg)^{-2.1} \bigg({t \over 1{\rm~day}}\bigg)^\delta~,
\label{eqa5}
\end{equation}
\begin{equation}
\tau_{\rm CSM}^{\rm clumps} = K_3 \bigg({\nu \over 5{\rm~GHz}}\bigg)^{-2.1} \bigg({t \over 1{\rm~day}}\bigg)^{\delta'}~,
\label{eqa6}
\end{equation}
and
\begin{equation}
\tau_{\rm SSA} = K_5 \bigg({\nu \over 5{\rm~GHz}}\bigg)^{\alpha-2.5} \bigg({t \over 1{\rm~day}}\bigg)^{\delta''}~.
\label{eqa7}
\end{equation}
Here, $K_1$ represents the unabsorbed flux density in mJy, and $K_2$, $K_3$, and $K_5$ are the optical depths for attenuation by a homogeneous absorbing CSM, a clumpy or filamentary CSM, and internal SSA, respectively, at day one after outburst and for $\nu=5$~GHz. The parameters $\delta$, $\delta'$, and $\delta''$ describe the time dependance of the optical depths $\tau_{\rm CSM}^{\rm homog}$, $\tau_{\rm CSM}^{\rm clumps}$, and $\tau_{\rm SSA}$. The parameters $\alpha$ and $\beta$ are the spectral index and rate of decline, respectively, in the optically thin phase. The model thus contains nine free parameters to be determined from fits to the data. The best-fit parameters can then be related to physical quantities in the system (see, e.g., Sect.~2.2 for the presupernova mass-loss rate). Other attenuation processes were taken into account for some SNe, e.g. FFA by an ionized medium on the light of sight far from the SN progenitor, but there are not important for SN~1993J (Weiler et al. \cite{wei07}). 

We show here that an estimate of the mean magnetic field strength in the radio emission region, $\langle B \rangle$, can be obtained from a comparison of this parametric model with the SSA model of Chevalier (\cite{che98}). In the latest formalism, the synchrotron emission is produced by relativistic electrons having a power-law energy distribution: $N_e(E)=N_0E^{-\gamma}$, where the spectral index $\gamma$ is independent of time. The emitting  spherical shell is approximated by a planar region in the plane of the sky with an area $\pi R^2$, where $R$ is the radius. In the absence of FFA by an external medium, the observable flux density can be written as
\begin{equation}
F_\nu = S_\nu  [1 - \exp(-\tau_{\rm SSA})]~,
\label{eqa8}
\end{equation}
with 
\begin{equation}
S_\nu = {\pi R^2 \over D^2} {c_5(\gamma) c_9(\gamma) \over c_6(\gamma) c'_9(\gamma)} \langle B \rangle^{-1/2} \bigg({\nu \over 2c_1}\bigg)^{5/2}~,
\label{eqa9}
\end{equation}
where $D$ is the source distance, the constant $c_1=6.26 \times 10^{18}$ in cgs units, $c_5(\gamma)$, $c_6(\gamma)$, and $c_9(\gamma)$ are tabulated as a function of $\gamma$ by Pacholczyk (\cite{pac70}) and 
\begin{equation}
c'_9(\gamma)={\sqrt{\pi} \over 2} {\Gamma[(\gamma+6)/4]\over\Gamma[(\gamma+8)/4]}~,
\label{eqa10}
\end{equation}
$\Gamma$ being the gamma function. The factors $c_9(\gamma)$ and $c'_9(\gamma)$ arise from the averaging of the synchrotron emission and absorption coefficients, respectively, over the particle pitch angle $\phi$ (Longair \cite{lon94})\footnote{These quantities were not taken into account by Chevalier (\cite{che98}), who made the approximation of replacing $\langle B \rangle \sin \phi$ by $\langle B \rangle$.}. We note that 
\begin{equation}
{c_5(\gamma) c_9(\gamma) \over c_6(\gamma) c'_9(\gamma)} = 3.49 \times 10^{17} c(\gamma)~~{\rm (cgs~units)},
\label{eqa11}
\end{equation}
where the expression for $c(\gamma)$ can be found in Fransson \& Bj\"ornsson (\cite{fra98}, Eq.~29). Calculated values of $c(\gamma)$ are given in Table~\ref{taba1} for seven values of $\gamma$. By equating the flux density from Eq.~(\ref{eqa8}) with the one obtained from the model of Weiler et al. (Eq.~\ref{eqa1}) for $A_{\rm CSM}^{\rm homog} = A_{\rm CSM}^{\rm clumps}=1$, we get
\begin{eqnarray}
\langle B \rangle & = & 0.135 c(\gamma)^2 \bigg({R_0 \over 10^{14} {\rm~cm}}\bigg)^4
\bigg({D \over 1 {\rm~Mpc}}\bigg)^{-4} \bigg({K_5 \over K_1}\bigg)^2 \nonumber \\
& \times & \bigg({t \over 1{\rm~day}}\bigg)^{4m-2(\beta-\delta'')}~~{\rm G}, 
\label{eqa12}
\end{eqnarray}
where we have identified the radius $R$ with the forward shock radius $R_s=R_0 (t/1~{\rm day})^m$. 

\begin{table}
\caption{Parameter $c(\gamma)$ for use in Eq.~(\ref{eqa12})}
\label{taba1}
\centering
\begin{tabular}{lccccccc}
\hline \hline
$\gamma$ & 1.5 & 2  & 2.5 & 3 & 3.5 & 4 & 4.5 \\
\hline
$c(\gamma)$ & 0.724 & 0.492 & 0.367 & 0.289 & 0.236 & 0.197 & 0.169 \\
\hline
\end{tabular}
\end{table}

Equation~(\ref{eqa12}) can in principle provide an estimate of the mean magnetic field in the synchrotron-emitting region as a function of the fitted parameters $K_1$, $K_5$, $\beta$, and $\delta''$. But in the case of combined SSA and FFA, the result strongly depends on the FFA model, which is uncertain (Sect.~2.2). Thus, with the best-fit parameters $\beta=-0.73$ and $\delta''=-2.05$ obtained for SN~1993J by Weiler et al. (\cite{wei07}), one gets the unreasonable time dependence $\langle B \rangle \propto t^{0.68}$ (for $m=0.83$). We argue in Sect.~4.3 that magnetic field amplification by cosmic-ray streaming should lead to $\langle B \rangle \propto t^{-b}$ with $b \approx 1$, as long as the shock is not strongly modified by the back pressure from the energetic ions. Thus, we expect the relation
\begin{equation}
\beta-\delta''\approx {4m+1 \over 2} 
\label{eqa13}
\end{equation}
to hold in most of the cases. Using this constraint in the fitting procedure could allow to better determine the physical parameters of radio SNe, in particular the mass loss rate of the progenitor star. 

\section{Radiative transfer calculations}

The synchrotron emission coefficient (in erg cm$^{-3}$ s$^{-1}$ sr$^{-1}$ Hz$^{-1}$) averaged over the pitch angle is given as a function of radius by (see, e.g., Pacholczyk \cite{pac70}) 
\begin{equation}
\epsilon_\nu(R)= {1 \over 8\pi} \int_{0}^{\pi}  \sin \phi d\phi \int_{0}^{\infty} p_\nu(E,\phi,R) N_e(E,R) dE
\label{eqb1}
\end{equation}
and the absorption coefficient (cm$^{-1}$) by
\begin{eqnarray}
\kappa_\nu(R) & = & -{c^2 \over 8 \pi \nu^2} \int_{0}^{\pi}  \sin \phi d\phi \int_{0}^{\infty} p_\nu(E,\phi,R) E^2 \nonumber \\
& \times & {\partial \over \partial E} \bigg({N_e(E,R) \over E^2}\bigg) dE~, 
\label{eqb2}
\end{eqnarray}
where $p_\nu(E,\phi,R)$ is the total emitted power per frequency per electron given in cgs units by the well-known formula
\begin{equation}
p_\nu(E,\phi,R)={\sqrt{3} e^3 B(R) \sin \phi \over m_e c^2} {\nu \over \nu_c} \int_{{\nu / \nu_c}}^{\infty} K_{5/3} (\eta) d\eta~.
\label{eqb3}
\end{equation}
Here, $-e$ is the electron charge, $K_{5/3}$ the modified Bessel function, and 
\begin{equation}
\nu_c = \frac{3 e}{4 \pi m_e^3 c^5} B(R) E^2 \sin \phi 
\label{eqb4}
\end{equation}
the critical frequency. In the computer code I have developed, the emission coefficient is calculated by numerical integration of Eq.~(\ref{eqb1}) using the tabulation of Pacholczyk (\cite{pac70}) for the integral of the modified Bessel function. The absorption coefficient is obtained by approximating the electron energy distribution by a power law $N_e(E_c,R)=N_0(R)E_c^{-\gamma}$ near the characteristic energy 
\begin{equation}
E_c = \bigg({4 \pi m_e^3 c^5 \nu \over 3 e B(R) \sin \phi}\bigg)^{1/2}~.
\label{eqb5}
\end{equation}
The absorption coefficient can then be simplified to (Pacholczyk \cite{pac70}; Longair \cite{lon94})
\begin{equation}
\kappa_\nu(R)=c_6(\gamma) c'_9(\gamma) N_0(R) B(R)^{(\gamma+2)/2} \bigg({\nu \over 2c_1}\bigg)^{-(\gamma+4)/2}. 
\label{eqb6}
\end{equation}
This approximation is used to speed up the computation time. Its accuracy has been checked by solving numerically Eq.~(\ref{eqb2}). 

The brightness profile, $I_\nu(\rho)$, where $\rho$ is the distance to the center of the disk resulting from the projection of the SNR onto the plane of the sky, is calculated from the source function $S_\nu \equiv \epsilon_\nu / \kappa_\nu$ by solving the standard transfer equation (Rybicki \& Lightman \cite{ryb79}):
\begin{equation}
I_\nu(\rho) = \int_{0}^{\tau_\nu} e^{-(\tau_\nu-\tau'_\nu)} S_\nu(\tau'_\nu) d\tau'_\nu~,
\label{eqb7}
\end{equation}
where the optical depths to SSA are defined as
\begin{eqnarray}
\tau_\nu & = & \int_{-\ell \times H(\rho-\rho_{\rm abs})}^{\ell} \kappa_\nu(R) dz~, \\
\tau'_\nu & = & \int_{-\ell \times H(\rho-\rho_{\rm abs})}^{z} \kappa_\nu(R) dz~, \label{eqb8}
\end{eqnarray}
with $\ell=\sqrt{R_s^2-\rho^2}$ and $z=\sqrt{R^2-\rho^2}$. The term $H(\rho-\rho_{\rm abs})$ accounts for attenuation of the emission from the side of the shell moving away from us by FFA in the interior of the SNR (e.g. Chevalier \cite{che82b}); here $H$ denotes the Heaviside step-function ($H(x)=0$ for $x<0$, $H(x)=1$ for $x \ge 0$). For simplicity, this absorption is modeled by a completely opaque disk of radius $\rho_{\rm abs}$ in the center of the remnant. Other absorption models were studied by Bietenholz et al. (\cite{bie03}). We note that FFA of the radio waves in the presupernova wind is not taken into account in Eq.~(\ref{eqb7}). In the limit of negligible SSA (i.e. optically-thin emission) Eq.~(\ref{eqb7}) reduces to
\begin{equation}
I_\nu(\rho) = \int_{-\ell \times H(\rho-\rho_{\rm abs})}^{\ell} \epsilon_\nu(R) dz~.
\label{eqb9}
\end{equation}

Finally, the total flux density is given by
\begin{equation}
F_\nu = {2\pi \over D^2} \int_{0}^{R_s} \rho I_\nu(\rho) d\rho~.
\label{eqb10}
\end{equation}
This quantity is multiplied by the attenuation factor $A_{\rm CSM}^{\rm clumps}$ (Eq.~\ref{eqa3}) for the modeling of the radio emission at early epochs when external FFA is important. 

\end{document}